\documentclass[sigplan,nonacm]{acmart}
\pdfoutput=1

\makeatletter                   
\def\mdseries@tt{m}             
\renewcommand\@formatdoi[1]{\ignorespaces} 
\let\@authorsaddresses\@empty 
\makeatother                    
\usepackage[plain]{fancyref}
\usepackage[draft=true]{minted} 
\usepackage{color}
\usepackage{hyperref}           
\hypersetup{
    colorlinks=true,
    linkcolor=blue,
    filecolor=red,
    urlcolor=magenta,
    breaklinks=true,            
}

\usepackage{mathpartir}
\usepackage{amsmath}
\usepackage{xspace}
\usepackage{subfiles}
\usepackage{float}
\floatstyle{boxed}
\restylefloat{figure}

\newif\ifanon
\makeatletter
\if@ACM@anonymous\anontrue\else\anonfalse\fi
\makeatother

\ifanon
\newcommand{\aquarium}{Aviary\xspace}

\newcommand{\ale}{Eagle\xspace}
\newcommand{\casp}{Kiwi\xspace}

\newcommand{\synth}{Hatch\xspace}
\else
\newcommand{\aquarium}{A\-quar\-ium\xspace}

\newcommand{\ale}{Ale\-wife\xspace}
\newcommand{\casp}{Cass\-io\-pea\xspace}

\newcommand{\synth}{Capstan\xspace}
\newcommand{\lowering}{low\-er\-ing\xspace}
\newcommand{\lowerings}{low\-er\-ings\xspace}
\fi

\usepackage{balance} 
\setcopyright{none} 
\acmDOI{} 
\acmISBN{} 
\acmYear{2019} 
\copyrightyear{2019} 

\author{David A. Holland}
\affiliation{Harvard University}
\email{dholland@eecs.harvard.edu}
\author{Jingmei Hu}
\affiliation{Harvard University}
\email{jingmei\_hu@g.harvard.edu}
\author{Ming Kawaguchi}
\affiliation{Harvard University}
\email{ming@seas.harvard.edu}
\author{Eric Lu}
\affiliation{Harvard University}
\email{ericlu01@g.harvard.edu}
\author{Stephen Chong}
\affiliation{Harvard University}
\email{chong@seas.harvard.edu}
\author{Margo I. Seltzer}
\affiliation{University of British Columbia}
\email{mseltzer@cs.ubc.ca}

\begin{document}
\sloppy

\title{\aquarium: \casp and \ale Languages \\
{\Large (Prerelease Version of 20220108)}
}

\begin{abstract}
This technical report describes two of the domain specific languages
used in the \aquarium kernel code synthesis project.
It presents the language cores in terms of abstract syntax.
\casp is a machine description language for describing the
semantics of processor instruction sets.
\ale is a specification language that can be used to write
machine-independent specifications for assembly-level instruction
blocks.
An \ale specification can be used to verify and synthesize code for
any machine described in \casp, given a machine-specific
translation for abstractions used in the specification.
This article does not include an introduction to either the \aquarium
system or the use of the languages.
In addition to this version of the article being a draft, the \aquarium
project and the languages are work in progress.
This article cannot currently be considered either final or complete.
\end{abstract}

\maketitle
\thispagestyle{empty}


 


\newcommand{\NEW}[1]{\textcolor{cyan}{#1}}
\newcommand{\NEWW}[1]{\textcolor{blue}{#1}}
\newcommand{\NNEW}[1]{\textcolor{green}{#1}}

\renewcommand{\NEW}[1]{#1}
\renewcommand{\NEWW}[1]{#1}
\renewcommand{\NNEW}[1]{#1}

\newcommand{\XXX}[1]{\textcolor{red}{#1}}

\newcommand{\ensm}{\XXX{WRONG}}
\newcommand{\ttid}{\XXX{WRONG}}
\newcommand{\ttmemid}{\XXX{WRONG}}
\newcommand{\ttmemidi}{\XXX{WRONG}}
\newcommand{\bvtyp}{\XXX{WRONG}}
\newcommand{\bvaltyp}{\XXX{WRONG}}
\newcommand{\bvltyp}{\XXX{WRONG}}
\newcommand{\ttC}{\XXX{WRONG}}
\newcommand{\ttN}{\XXX{WRONG}}
\newcommand{\ttloc}{\XXX{WRONG}}
\newcommand{\ttmlen}{\XXX{WRONG}}
\newcommand{\ttbit}{\XXX{WRONG}}
\newcommand{\ttwptr}{\XXX{WRONG}}
\newcommand{\ttwvec}{\XXX{WRONG}}
\newcommand{\ttbool}{\XXX{WRONG}}
\newcommand{\ttint}{\XXX{WRONG}}
\newcommand{\ttinparen}{\XXX{WRONG}}
\newcommand{\atyp}{\XXX{WRONG}}
\newcommand{\memtyp}{\XXX{WRONG}}
\newcommand{\labeltyp}{\XXX{WRONG}}
\newcommand{\nbitmemtyp}{\XXX{WRONG}}
\newcommand{\ptrform}{\XXX{WRONG}}
\newcommand{\bnfeq}{\XXX{WRONG}}
\newcommand{\bnfor}{\XXX{WRONG}}
\newcommand{\aprimitivetyp}{\XXX{WRONG}}

%


\newcommand{\mynote}[1]{\text{{\begin{tiny}\textbf{MK: }#1\end{tiny}}}}

\newcommand{\mypara}[1]{\textbf{#1.}}

\newcommand{\ie}{\textit{i.e.,}}
\newcommand{\eg}{\textit{e.g.,}}
\newcommand{\etc}{\textit{etc.}}

\renewcommand{\ensm}[1]{\ensuremath{#1}}

\newcommand{\eps}{\ensm{\epsilon}}

\newcommand{\holds}{\ensm{\vdash}}
\newcommand{\sholds}{\ensm{\models}}

\newcommand{\tsp}{\thinspace}

\newcommand{\bitnot}{\ensm{\mathtt{bnot}}\xspace}
\newcommand{\bitadd}{\ensm{\mathtt{b+}}\xspace}
\newcommand{\bitsub}{\ensm{\mathtt{b-}}\xspace}
\newcommand{\bitmul}{\ensm{\mathtt{b*}}\xspace}
\newcommand{\bitdiv}{\ensm{\mathtt{b/}}\xspace}
\newcommand{\bitlt}{\ensm{\mathtt{\mathop{b\mathord{<}}}}\xspace}
\newcommand{\bitle}{\ensm{\mathtt{\mathop{b\mathord{<=}}}}\xspace}
\newcommand{\bitgt}{\ensm{\mathtt{\mathop{b\mathord{>}}}}\xspace}
\newcommand{\bitge}{\ensm{\mathtt{\mathop{b\mathord{>=}}}}\xspace}
\newcommand{\bitsle}{\ensm{\mathtt{\mathop{bs\mathord{<=}}}}\xspace}
\newcommand{\bitsge}{\ensm{\mathtt{\mathop{bs\mathord{>=}}}}\xspace}
\newcommand{\bitslt}{\ensm{\mathtt{\mathop{bs\mathord{<}}}}\xspace}
\newcommand{\bitsgt}{\ensm{\mathtt{\mathop{bs\mathord{>}}}}\xspace}
\newcommand{\bitand}{\ensm{\mathtt{band}}\xspace}
\newcommand{\bitor}{\ensm{\mathtt{bor}}\xspace}
\newcommand{\bitxor}{\ensm{\mathtt{bxor}}\xspace}
\newcommand{\intdiv}{\ensm{\mathtt{div}}\xspace}
\newcommand{\setdiff}{\ensm{\mathtt{s-}}\xspace}
\newcommand{\xor}{\mathbin{\oplus}}

\newcommand{\defeq}{\ensm{=}\xspace}
\newcommand{\ttinclude}{\ensm{\mathtt{include}}\xspace}
\newcommand{\ttskip}{\ensm{\mathtt{skip}}\xspace}
\newcommand{\ttsemi}{\ensm{\mathit{;}}\xspace}
\newcommand{\ttcol}{\ensm{\mathit{:}}\xspace}
\newcommand{\ttcomma}{\ensm{\mathit{,}\ }\xspace}
\newcommand{\ttdot}{\ensm{\mathit{.}}\xspace}
\newcommand{\tttrue}{\ensm{\mathtt{true}}\xspace}
\newcommand{\ttfalse}{\ensm{\mathtt{false}}\xspace}
\newcommand{\tterr}{\ensm{\mathit{err}}\xspace}
\newcommand{\ttassert}{\ensm{\mathtt{assert}}\xspace}
\newcommand{\ttassertfalse}{\ensm{\mathtt{fail}}\xspace}
\newcommand{\ttwrong}{\ensm{\mathit{wrong}}\xspace}
\newcommand{\ttinvalid}{\ensm{\mathit{invalid}}\xspace}
\newcommand{\ttvalid}{\ensm{\mathit{valid}}\xspace}
\newcommand{\itbinop}{\ensm{\mathit{binop}}\xspace}
\newcommand{\itsetop}{\ensm{\mathit{setop}}\xspace}
\newcommand{\itboolop}{\ensm{\mathit{boolop}}\xspace}
\newcommand{\itbinrel}{\ensm{\mathit{binrel}}\xspace}
\newcommand{\itbintop}{\ensm{\mathit{bintop}}\xspace}
\newcommand{\itbvecop}{\ensm{\mathit{bvecop}}\xspace}
\newcommand{\itunop}{\ensm{\mathit{unop}}\xspace}
\newcommand{\itatomic}{\ensm{\mathit{atomic}}\xspace}
\newcommand{\ttvec}{\ensm{\mathit{vec}}\xspace}
\renewcommand{\ttid}{\ensm{\mathit{id}}\xspace}
\newcommand{\ttxmoduleid}{\ensm{\mathit{{x}_{module}}}\xspace}
\newcommand{\ttxblockid}{\ensm{\mathit{{x}_{block}}}\xspace}
\newcommand{\ttfuncid}{\ensm{\mathit{x_{func}}}\xspace}
\newcommand{\ttprocid}{\ensm{\mathit{x_{proc}}}\xspace}
\newcommand{\ttxop}{\ensm{\mathit{x_{op}}}\xspace}
\newcommand{\ttregid}{\ensm{\mathit{x_{reg}}}\xspace}
\newcommand{\ttregidi}{\ensm{\mathit{x_{{\mathit{reg}}_i}}}\xspace}
\renewcommand{\ttmemid}{\ensm{\mathit{x_{mem}}}\xspace}
\renewcommand{\ttmemidi}{\ensm{\mathit{x_{{\mathit{mem}}_i}}}\xspace}
\newcommand{\ttlabelid}{\ensm{\mathit{x_{label}}}\xspace}
\newcommand{\tttypeid}{\ensm{\mathit{x_{\tau}}}\xspace}
\newcommand{\ttextid}{\ensm{\mathit{x_{ext}}}\xspace}
\newcommand{\ttidents}{\ensm{\mathit{identifiers}}\xspace}
\newcommand{\ttunit}{\ensm{\mathit{()}}\xspace}
\newcommand{\ttargs}{\ensm{\mathit{args}}\xspace}
\newcommand{\ttstr}{\ensm{\mathit{string}}\xspace}
\newcommand{\ttget}{\ensm{\mathtt{get}}\xspace}
\newcommand{\ttassign}{\ensm{\mathtt{assign}}\xspace}
\newcommand{\ttindex}{\ensm{\mathit{index}}\xspace}
\newcommand{\ttusing}{\ensm{\mathit{with}}\xspace}
\newcommand{\ttread}{\ensm{\mathtt{read}}\xspace}
\newcommand{\ttfetch}{\ensm{\mathtt{fetch}}\xspace}
\newcommand{\ttwrite}{\ensm{\mathtt{write}}\xspace}
\newcommand{\ttstore}{\ensm{\mathtt{store}}\xspace}
\newcommand{\ttat}{\ensm{{\mathit{\small @}}}\xspace}
\newcommand{\ttdecl}{\ensm{\mathit{decl}\xspace}}
\newcommand{\ttdecls}{\ensm{\mathit{decls}\xspace}}
\newcommand{\ttlet}{\ensm{\mathtt{let}}\xspace}
\newcommand{\ttletstate}{\ensm{\mathtt{letstate}}\xspace}
\newcommand{\ttdef}{\ensm{\mathtt{def}}\xspace}
\newcommand{\ttproc}{\ensm{\mathtt{proc}}\xspace}
\newcommand{\itlet}{\ensm{\mathit{let}}\xspace}
\newcommand{\itlets}{\ensm{\mathit{lets}}\xspace}
\newcommand{\tttype}{\ensm{\mathtt{type}}\xspace}
\newcommand{\ttstate}{\ensm{\mathtt{state}}\xspace}
\newcommand{\ttcontrol}{\ensm{\mathtt{control}}\xspace}
\newcommand{\ttdontgate}{\ensm{\mathtt{dontgate}}\xspace}
\newcommand{\ttastate}{\ensm{\mathtt{state}}\xspace}
\newcommand{\ttdefop}{\ensm{\mathtt{defop}}\xspace}
\newcommand{\ttend}{\ensm{\mathtt{end}}\xspace}
\newcommand{\ttmach}{\ensm{\mathit{machine}}\xspace}
\newcommand{\ttprog}{\ensm{\mathit{program}}\xspace}
\newcommand{\tttxt}{\ensm{\mathtt{txt}}\xspace}
\newcommand{\tthex}{\ensm{\mathtt{hex}}\xspace}
\newcommand{\ttdec}{\ensm{\mathtt{dec}}\xspace}
\newcommand{\ttbin}{\ensm{\mathtt{bin}}\xspace}
\newcommand{\ttlbl}{\ensm{\mathtt{lbl}}\xspace}
\newcommand{\ttsem}{\ensm{\mathtt{sem}}\xspace}
\newcommand{\ttinit}{\ensm{\mathit{init}}\xspace}
\newcommand{\ttfunc}{\ensm{\mathit{function}}\xspace}
\newcommand{\ttval}{\ensm{\mathit{val}}\xspace}
\newcommand{\ttvalue}{\ensm{\mathit{Value}}\xspace}
\newcommand{\ttcast}{\ensm{\mathtt{coerce}\tsp\mathtt{as}}\xspace}
\newcommand{\tthavoc}{\ensm{\mathit{nondet()}\xspace}}
\newcommand{\ttaddress}{\ensm{\mathit{Addr}\xspace}}
\newcommand{\ttis}{\ensm{\mathtt{is}\xspace}}
\newcommand{\ttcptr}{\ensm{\mathtt{ptr}}\xspace}
\newcommand{\ttcPtr}{\ensm{\mathtt{Ptr}}\xspace}
\newcommand{\tteptr}{\ensm{\mathtt{ptr}}\xspace}
\newcommand{\ttvars}{\ensm{\mathit{vars}}\xspace}
\newcommand{\ttvals}{\ensm{\mathit{values}}\xspace}
\newcommand{\ttglobals}{\ensm{\mathit{globals}}\xspace}
\newcommand{\ttterm}{\ensm{T}\xspace}
\newcommand{\ttor}{\ensm{\mathit{or}}\xspace}
\newcommand{\ttfresh}{\ensm{\mathit{fresh}}\xspace}
\newcommand{\ttmod}{\ensm{\mathtt{mod}}\xspace}
\newcommand{\ttthen}{\ensm{\rhd}\xspace}

\newcommand{\exprfail}{\ttassertfalse}

\newcommand{\ttbaseof}{\ensm{\mathtt{base}}}
\newcommand{\ttoffsetof}{\ensm{\mathtt{offset}}}
\newcommand{\ttasbitv}{\ensm{\mathit{(as\ttbit)}}}

\newcommand{\clholds}{\ensm{\vdash_{pred}}\tsp}
\newcommand{\clog}{\ensm{\phi}}
\newcommand{\calog}{\ensm{\clog_\forall}}
\newcommand{\celog}{\ensm{\clog_\exists}}
\newcommand{\cor}{\ensm{\vee}}
\newcommand{\cand}{\ensm{\wedge}}
\newcommand{\cimplies}{\ensm{\rightarrow}}
\newcommand{\ceq}{\ensm{=}}

\renewcommand{\bvtyp}[1]{\ensm{#1\ \ttbit}}
\newcommand{\cptrtyp}[1]{\ensm{\bvtyp{#1}}}
\renewcommand{\bvaltyp}[1]{\ensm{\bvtyp{#1}}}
\renewcommand{\bvltyp}[1]{\ensm{#1\ \ttloc}}
\newcommand{\bvstyp}[1]{\ensm{#1\ \ttloc\ \ttset}}


\renewcommand{\ttC}{\ensm{\mathit{C}}\xspace}
\newcommand{\ttR}{\ensm{\mathit{R}}\xspace}
\newcommand{\ttInt}{\ensm{\mathit{Int}}\xspace}
\newcommand{\ttVec}{\ensm{\mathit{Vec}}\xspace}

\newcommand{\ttmap}{\ensm{\mathit{map}}\xspace}

\newcommand{\cfuntyp}{\atyp_{\mathit{func}}}
\newcommand{\cbasetyp}{\atyp_{\mathit{base}}}
\newcommand{\cbasetypi}{\atyp_{{\mathit{base}}_i}}
\newcommand{\cbasetypone}{\ensm{\atyp_{{\mathit{base}}_1}}}
\newcommand{\cbasetyptwo}{\ensm{\atyp_{{\mathit{base}}_2}}}
\newcommand{\cregtyp}{\atyp_{\mathit{reg}}}
\newcommand{\cregstyp}{\atyp_{\mathit{regs}}}
\newcommand{\cmemtyp}{\atyp_{\mathit{mem}}}
\newcommand{\clbltyp}{\atyp_{\mathit{label}}}
\newcommand{\cstatetyp}{\atyp_{\mathit{state}}}

\newcommand{\ttmem}{\ensm{\mathtt{mem}}\xspace}
\newcommand{\ttlabel}{\ensm{\mathtt{label}}\xspace}
\renewcommand{\ttloc}{\ensm{\mathtt{reg}}\xspace}

\newcommand{\ttlen}{\ensm{\mathit{len}}\xspace}
\renewcommand{\ttmlen}{\ensm{\mathtt{len}}\xspace}
\renewcommand{\ttbit}{\ensm{\mathtt{bit}}\xspace}
\newcommand{\ttref}{\ensm{\mathtt{ref}}\xspace}
\renewcommand{\ttwptr}{\ensm{\mathtt{ptr}}\xspace}
\renewcommand{\ttwvec}{\ensm{\mathtt{vec}}\xspace}
\newcommand{\ttBit}{\ensm{\mathtt{Bits}}\xspace}
\newcommand{\ttset}{\ensm{\mathtt{set}}\xspace}
\renewcommand{\ttbool}{\ensm{\mathtt{bool}}\xspace}
\renewcommand{\ttint}{\ensm{\mathtt{int}}\xspace}
\newcommand{\ttstring}{\ensm{\mathtt{string}}\xspace}
\newcommand{\ttn}{\ensm{\mathit{n}}\xspace}
\newcommand{\ttk}{\ensm{\mathit{k}}\xspace}
\newcommand{\ttK}{\ensm{\mathit{K}}\xspace}
\newcommand{\tti}{\ensm{\mathit{i}}\xspace}
\newcommand{\ttj}{\ensm{\mathit{j}}\xspace}
\newcommand{\ttb}{\ensm{\mathit{b}}\xspace}
\newcommand{\ttv}{\ensm{\mathit{v}}\xspace}
\newcommand{\tildev}{\ensm{\mathit{\tilde{v}}}\xspace}
\newcommand{\ttr}{\ensm{\mathit{r}}\xspace}
\newcommand{\tte}{\ensm{\mathit{e}}\xspace}
\newcommand{\ttS}{\ensm{\mathit{S}}\xspace}
\renewcommand{\ttN}{\ensm{\mathit{N}}\xspace}
\newcommand{\ttx}{\ensm{\mathit{x}}\xspace}
\newcommand{\tty}{\ensm{\mathit{y}}\xspace}
\newcommand{\ttz}{\ensm{\mathit{z}}\xspace}
\newcommand{\ttl}{\ensm{\mathit{l}}\xspace}
\newcommand{\ttT}{\ensm{\mathit{T}}\xspace}
\newcommand{\ttm}{\ensm{\mathit{m}}\xspace}
\newcommand{\ttM}{\ensm{\mathit{M}}\xspace}
\newcommand{\ttV}{\ensm{\mathit{V}}\xspace}

\newcommand{\ttconst}{\ensm{const}\xspace}

\newcommand{\globalv}[1]{\ensm{#1}\xspace}

\newcommand{\ttchoice}{\ensm{\mathit{\ ?\ }}\xspace}
\newcommand{\ttinbracket}[1]{\ensm{\mathit{[}#1\mathit{]}}\xspace}
\renewcommand{\ttinparen}[1]{\ensm{\mathit{(}#1\mathit{)}}\xspace}
\newcommand{\ttinbrace}[1]{\ensm{\mathit{\{}#1\mathit{\}}}\xspace}
\newcommand{\inhline}[1]{\ensm{\vert #1\vert}\xspace}

\newcommand{\bitconstprefix}[1]{\ensm{\mathtt{0b}#1}}
\newcommand{\bitconst}{\ensm{\bitconstprefix{\ttC}}}
\newcommand{\hexconst}{\ensm{\mathtt{0x}\ttC}}

\newcommand{\sizeof}[1]{\ensm{\ttsizeof(#1)}\xspace}

\newcommand{\setlit}{\ensm{\ttinbrace{\ttx_1, \ldots, \ttx_k}}\xspace}
\newcommand{\intsetlit}{\ensm{\ttinbrace{\ttC, \ldots \ttC}}}
\newcommand{\ttsetemp}{\ensm{\ttinbrace{}}\xspace}
\newcommand{\sizeofset}[1]{\ensm{\|{#1}\|}\xspace}
\newcommand{\setmemberof}[2]{\ensm{#1\tsp{\in}#2}}
\newcommand{\setcomplement}[1]{\overline{\ensm{#1}^c}}

\newcommand{\program}{\ensm{\ttprog}}
\newcommand{\machine}{\ensm{\ttmach}}
\newcommand{\decls}{\ensm{\ttdecls}}
\newcommand{\declim}{\ensm{\ttsemi\ }}
\newcommand{\decl}{\ensm{\ttdecl}}
\newcommand{\defops}{\ensm{\mathit{defops}}\xspace}
\newcommand{\defop}{\ensm{\mathit{defop}}\xspace}
\newcommand{\defeffectfunc}{\ensm{\ttdefsproc}}

\newcommand{\meminit}[1]{\ensm{\ttinit\ \ttindex\ \tti\ \ttusing\ (#1)}} 

\newcommand{\bfe}{\ensm{\mathbf{e}}}
\newcommand{\bferr}{\ensm{\mathtt{crash}}}

\newcommand{\subst}[2]{\ensm{[#1\mathbf{\setminus} #2]}}

\newcommand{\lderef}[1]{\regderef}
\newcommand{\memread}[1]{\ttfetch(#1,\ttC)}
\newcommand{\fetch}[2]{\ttfetch(#1,#2)}
\newcommand{\absfetch}[1]{\ttfetch(#1)}
\newcommand{\regderef}[1]{\:^*#1}
\newcommand{\dholds}{\ensm{\vDash}}
\newcommand{\tholds}{\ensm{\vdash}}
\newcommand{\twfholds}{\ensm{\vdash_{\texttt{wf}}}}
\newcommand{\sqright}{\rightsquigarrow}
\newcommand{\sqrightm}[1]{\rightsquigarrow_{#1}}
\newcommand{\assign}[2]{\ensm{\ttstore(#1) \leftarrow #2}}
\newcommand{\callstmt}[2]{\ensm{\ttcall\ #1\ #2}}

\newcommand{\sseq}[1]{\ensm{\overline{#1}}}
\newcommand{\pseq}[2]{\ensm{\widehat{#1\ttcol #2}}}

\renewcommand{\atyp}{\ensm{\tau}\xspace}
\newcommand{\funtyp}[2]{\ensm{#1 \rightarrow #2}\xspace}
\newcommand{\proctyp}[1]{\ensm{#1 \rightarrow ()}\xspace}

\newcommand{\hastyp}[2]{\ensm{#1\ \mathit{:}\ #2}}
\newcommand{\oftyp}[2]{\ensm{#1\ \mathit{:}\ #2}}


\renewcommand{\memtyp}[3]{\ensm{#1\ \ttbit\ #2\ \ttmlen\ #3\ \texttt{ref}}}
\renewcommand{\labeltyp}[1]{\ensm{#1\ \ttlabel}}
\renewcommand{\nbitmemtyp}[1]{\memtyp{\_}{\_}{#1}}

\newcommand{\cptr}[2]{\ensm{\ttcptr(#1,#2)}}
\newcommand{\eptr}[2]{\ensm{\tteptr(#1,#2)}}
\newcommand{\ptrtyp}{\ensm{\ttaptr}}
\renewcommand{\ptrform}[2]{\ensm{\ttinparen{#1 \ttcomma #2}}}
\newcommand{\lblform}[1]{\ensm{\ttinbrace{#1}}}


\newcommand{\isptr}[1]{\ensm{\ttis\ttcptr(#1)}}
\newcommand{\addrof}[1]{\ensm{\ttbaseof(#1)}}
\newcommand{\offsetof}[1]{\ensm{\ttoffsetof(#1)}}
\newcommand{\bvcast}[1]{\ensm{\ttasbitv(#1)}}

\newcommand{\tyvar}{\ensm{\alpha}}

\newcommand{\bmapsto}[2]{\ensm{\mathit{[}{#1}\rightarrowtail{#2}}\mathit{]}}

\newcommand{\fapp}[2]{\ensm{#1\ (#2)}}
\newcommand{\papp}[2]{\ensm{#1\ (#2)}}

\newcommand{\offset}[2]{\ensm{#1\mathit{@}#2}}

\newcommand{\irule}[2]{\ensm{\inferrule{#1}{#2}}}
\newcommand{\actrans}[1]{\ensm{\mathcal{AC}\llbracket#1\rrbracket}}
\newcommand{\actrule}[1]{\ensm{\typenv, \kenv \holds \actrans{#1}}}
\newcommand{\acrule}[1]{\ensm{\typenv, \kenv \holds \actrans{#1}}}
\newcommand{\acruleCC}[1]{\ensm{\typenvCC, \kenv \holds \actrans{#1}}}
\newcommand{\actyptrans}[1]{\ensm{\mathcal{AC}\llbracket#1\rrbracket}}
\newcommand{\actyprule}[1]{\ensm{\typenv, \kenv \holds \actyptrans{#1}}}
\newcommand{\acsigma}[1]{\ensm{\Sigma(#1)}}
\newcommand{\acphi}[1]{\ensm{\Phi(#1)}}
\newcommand{\aclambda}[1]{\ensm{\venv(#1)}}
\newcommand{\acdelta}[1]{\ensm{\genv(#1)}}
\newcommand{\acgamma}[1]{\ensm{\tenv(#1)}}
\newcommand{\st}{\mathbf{.}}

\newcommand{\sITE}[3]{\ensm{\texttt{if}\tsp#1\tsp\texttt{then}\tsp#2\tsp\texttt{else}\tsp#3}}
\newcommand{\eITE}[3]{\ensm{#1\texttt{ ? }#2\texttt{ : }#3}}

\newcommand{\sLet}[4]{\ensm{\texttt{let}\tsp\oftyp{#1}{#2}= #3\tsp\texttt{in}\tsp#4}}
\newcommand{\eLet}[4]{\ensm{\texttt{let}\tsp\oftyp{#1}{#2}= #3\tsp\texttt{in}\tsp#4}}

\newcommand{\memidx}[1]{\:^*#1}
\newcommand{\sFor}[4]{\ensm{\texttt{for}\tsp#1\in(#2\ldots #3)\ \texttt{do}\tsp#4}}
\newcommand{\smAssign}[2]{\assign{#1}{#2}}
\newcommand{\slAssign}[2]{#1\tsp\ttcol= #2}

\newcommand{\ebranchto}[1]{\ensm{\texttt{branchto}}}
\newcommand{\sBRANCH}[1]{\ensm{\texttt{BRANCH}(#1)}}

\newcommand{\bitidx}[2]{\ensm{#1\ttinbracket{#2}}}
\newcommand{\bitfield}[3]{\ensm{#1\ttinbracket{#2\ttcomma #3}}}

\newcommand{\execmemidx}[2]{#1 (#2)}
\newcommand{\execmenvidx}[2]{#1 (#2)}

\renewcommand{\bnfeq}{\ensm{\Coloneqq}}
\renewcommand{\bnfor}{\ensm{\mid}}


\newcommand{\strconcat}[2]{\ensm{#1\ \texttt{++}\ #2}}
\newcommand{\strof}[1]{\ensm{#1\ttdot\tttxt}}
\newcommand{\hexof}[1]{\ensm{#1\ttdot\tthex}}
\newcommand{\decof}[1]{\ensm{#1\ttdot\ttdec}}
\newcommand{\binof}[1]{\ensm{#1\ttdot\ttbin}}
\newcommand{\lblof}[1]{\ensm{#1\ttdot\ttlbl}}
\newcommand{\strlit}{\texttt{"}$\ldots$\texttt{"}}


\newcommand{\genv}{\Delta} 
\newcommand{\tenv}{\Gamma} 
\newcommand{\venv}{\Lambda}
\newcommand{\env}{\ensm{\tenv}\xspace} 
\newcommand{\wfenv}{\ensm{\genv}}
\newcommand{\typenv}{\ensm{\genv, \tenv}}
\newcommand{\typenvAB}{\ensm{\genv, \tenv'}}
\newcommand{\typenvAC}{\ensm{\genv, \tenv''}}
\newcommand{\typenvBA}{\ensm{\genv', \tenv}}
\newcommand{\typenvBB}{\ensm{\genv', \tenv'}}
\newcommand{\typenvCC}{\ensm{\genv'', \tenv''}}
\newcommand{\subenv}{\ensm{\Sigma}}
\newcommand{\subenvZ}{\ensm{\Sigma_{\mathit{builtin}}}}
\newcommand{\subenvB}{\ensm{\Sigma'}}
\newcommand{\subenvC}{\ensm{\Sigma''}}

\newcommand{\alenv}{\ensm{\Sigma, \Phi}}
\newcommand{\alenvZZ}{\ensm{\Sigma_{\mathit{builtin}}, \Phi_{\mathit{builtin}}}}
\newcommand{\alenvAB}{\ensm{\Sigma, \Phi'}}
\newcommand{\alenvAC}{\ensm{\Sigma, \Phi''}}
\newcommand{\alenvBA}{\ensm{\Sigma', \Phi}}
\newcommand{\alenvBB}{\ensm{\Sigma', \Phi'}}
\newcommand{\alenvCC}{\ensm{\Sigma'', \Phi''}}
\newcommand{\denv}{\ensm{\tenv,\venv}\xspace} 

\newcommand{\subt}{<:}

\newcommand{\nextrulesmall}{\vskip 8pt}
\newcommand{\nextrule}{\vskip 15pt}
\newcommand{\nextclause}{\ensm{}}

\newcommand{\envgoesto}{\ensm{\mapsto}}
\newcommand{\goesto}{\ensm{\rightarrow}}
\newcommand{\dngoesto}{\ensm{\nrightarrow}}
\newcommand{\egoesto}{\ensm{\Downarrow}}
\newcommand{\sgoesto}{\ensm{\Downarrow}}
\newcommand{\opstenv}{\ensm{\Gamma}}
\newcommand{\opsmenv}{\ensm{\sigma}}
\newcommand{\opsrenv}{\ensm{\rho}}
\newcommand{\memgoesto}{\ensm{\rightarrow}}

\newcommand{\opsenv}{\ensm{\opsrenv, \opsmenv}\xspace}
\newcommand{\opsenvB}{\ensm{\opsrenv', \opsmenv'}\xspace}
\newcommand{\opsenvC}{\ensm{\opsrenv'', \opsmenv''}\xspace}
\newcommand{\opsenvafter}[2]{\ensm{\opsrenv_{#2},\opsmenv_{#1}}\xspace}
\newcommand{\oholds}{\ensm{\vdash}\xspace}

\newcommand{\osreduce}[4]{\osreducetenv{\venv}{#1}{#2}{#3}{#4}}
\newcommand{\osreduceafter}[4]{\osreducetenv{\venv'}{#1}{#2}{#3}{#4}}
\newcommand{\osreducetenv}[5]{\ensm{#1\oholds(#3,#2)\sgoesto(#5,#4,\NEW{\cdot})}}
\newcommand{\osreduceBR}[5]{\osreducetenvBR{\venv}{#1}{#2}{#3}{#4}{#5}}
\newcommand{\osreduceafterBR}[5]{\osreducetenvBR{\venv'}{#1}{#2}{#3}{#4}{#5}}
\newcommand{\osreducetenvBR}[6]{\ensm{#1\oholds(#3,#2)\sgoesto(#5,#4,\NEW{#6})}}

\newcommand{\psreduce}[3]{\preducetenv{\venv}{#1}{#2}{#3}}
\newcommand{\psreduceB}[3]{\preducetenv{\venv'}{#1}{#2}{#3}}
\newcommand{\preducetenv}[4]{\ensm{#1\oholds(#3,#2)\goesto(#4)}}

\newcommand{\psreducemulti}[3]{\preducetenvmulti{\venv}{#1}{#2}{#3}}
\newcommand{\psreducemultiB}[3]{\preducetenvmulti{\venv'}{#1}{#2}{#3}}
\newcommand{\preducetenvmulti}[4]{\ensm{#1\oholds(#3,#2)\goesto^*(#4)}}

\newcommand{\esloclookup}[1]{\ensm{\opsrenv(#1) = \ttv}}
\newcommand{\esmemlookup}[3]{\ensm{\opsmenv(#1, #2) = #3}}
\newcommand{\esred}[4]{\ensm{\venv#4\oholds(\opsenv,#1)#3#2}}
\newcommand{\esreduce}[2]{\esred{#1}{#2}{\egoesto}{}{}}
\newcommand{\esredafter}[4]{\ensm{\venv#4\oholds(\opsrenv', \opsmenv',#1)#3#2}}
\newcommand{\esreduceafter}[2]{\esredafter{#1}{#2}{\egoesto}{}{}}
\newcommand{\esredB}[4]{\ensm{\venv'#4\oholds(\opsenv,#1)#3#2}}
\newcommand{\esreduceB}[2]{\esredB{#1}{#2}{\egoesto}{}{}}
\newcommand{\esredafterB}[4]{\ensm{\venv'#4\oholds(\opsrenv', \opsmenv',#1)#3#2}}
\newcommand{\esreduceafterB}[2]{\esredafterB{#1}{#2}{\egoesto}{}{}}
\newcommand{\esreducevenv}[4]{\ensm{#1\oholds(#3,#2)\egoesto#4}}
\newcommand{\esdnreduce}[2]{\esred{#1}{#2}{\dngoesto}{}{}}


\newcommand{\rbox}[1]{\ensm{r}}
\newcommand{\rboxid}{\rbox{\globid}}
\newcommand{\membox}[1]{\ensm{m}}
\newcommand{\memboxid}{\membox{\globid}}

\newcommand{\SEMdeclZA}[1]{\ensm{  \venv_{\mathit{builtin}}, (\{\}, \{\}) \holds #1 \ttthen \venv }}
\newcommand{\SEMdeclAA}[1]{\ensm{  \venv, (\opsrenv, \opsmenv) \holds  #1  \ttthen \venv  }}
\newcommand{\SEMdeclAB}[1]{\ensm{  \venv, (\opsrenv, \opsmenv) \holds  #1  \ttthen \venv'  }}
\newcommand{\SEMdeclAC}[1]{\ensm{  \venv, (\opsrenv, \opsmenv) \holds  #1  \ttthen \venv''  }}
\newcommand{\SEMdeclBC}[1]{\ensm{  \venv', (\opsrenv, \opsmenv) \holds #1  \ttthen \venv''  }}
\newcommand{\SEMdeclAx}[2]{\ensm{  \venv, (\opsrenv, \opsmenv) \holds  #1  \ttthen \venv[#2]  }}
\newcommand{\SEMdeclZ}[1]{\ensm{  \holds, (\opsrenv, \opsmenv)  #1  \ttthen \venv_0  }}
\newcommand{\SEMdeclA}[1]{\ensm{  \holds, (\opsrenv, \opsmenv)  #1  \ttthen \venv  }}
\newcommand{\SEMdeclB}[1]{\ensm{  \holds, (\opsrenv, \opsmenv)  #1  \ttthen \venv'  }}
\newcommand{\SEMZZA}[1]{\ensm{  \venv_{\mathit{builtin}}, (\{\}, \{\}) \holds #1 \ttthen \venv }}
\newcommand{\SEMAAAA}[1]{\ensm{  \venv, (\opsrenv, \opsmenv) \holds  #1  \ttthen \venv, (\opsrenv, \opsmenv)  }}
\newcommand{\SEMAAB}[1]{\ensm{  \venv, (\opsrenv, \opsmenv) \holds  #1  \ttthen \venv' }}
\newcommand{\SEMAA}[1]{\ensm{  \holds  #1  \ttthen \venv, (\opsrenv, \opsmenv) }}
\newcommand{\SEMBA}[1]{\ensm{  \holds  #1  \ttthen \venv', (\opsrenv, \opsmenv) }}

\newcommand{\SEMmachZ}[1]{\ensm{  \holds  #1  \ttthen \venv_0  }}
\newcommand{\SEMmachA}[1]{\ensm{  \holds  #1  \ttthen \venv  }}
\newcommand{\SEMmachB}[1]{\ensm{  \holds  #1  \ttthen \venv'  }}
\newcommand{\SEMmachAB}[1]{\ensm{  \venv \holds  #1  \ttthen \venv'  }}

\newcommand{\ttae}{\ensm{\mathit{e}}}
\newcommand{\ttce}{\ensm{\mathit{c\text{-}e}}}
\newcommand{\italewifespec}{\ensm{\mathit{ale\text{-}spec}}}
\newcommand{\italewifebinding}{\ensm{\mathit{ale\text{-}binding}}}
\newcommand{\italedecls}{\ensm{\mathit{ale\text{-}decls}}}
\newcommand{\italedecl}{\ensm{\mathit{ale\text{-}decl}}}
\newcommand{\itaxioms}{\ensm{\mathit{axioms}}\xspace}
\newcommand{\itderivations}{\ensm{\mathit{derivations}}\xspace}
\newcommand{\itdeclarations}{\ensm{\mathit{declarations}}\xspace}
\newcommand{\itdecl}{\ensm{\mathit{decl}}\xspace}
\newcommand{\itdecls}{\ensm{\mathit{decls}}\xspace}
\newcommand{\itdefinitions}{\ensm{\mathit{definitions}}\xspace}
\newcommand{\itblocks}{\ensm{\mathit{blocks}}\xspace}
\newcommand{\ttaxiom}{\ensm{\mathit{axiom}}\xspace}
\newcommand{\ttBlock}{\ensm{\mathtt{Block}}\xspace}
\newcommand{\itblock}{\ensm{\mathit{block}}\xspace}
\newcommand{\ttblock}{\ensm{\mathtt{block}}\xspace}
\newcommand{\itspec}{\ensm{\mathit{spec}}\xspace}
\newcommand{\itinsts}{\ensm{\mathit{insts}}\xspace}
\newcommand{\itinst}{\ensm{\mathit{inst}}\xspace}
\newcommand{\ttspecdef}{\ensm{\mathit{define\text{-}spec}}\xspace}
\newcommand{\ttblockdef}{\ensm{\mathit{define\text{-}block}}\xspace}
\newcommand{\ttblockbind}{\ensm{\itblock\text{-}\itlet}\xspace}
\newcommand{\ttblockbinds}{\ensm{\itblock\text{-}\itlets}\xspace}
\newcommand{\itregion}{\ensm{\mathit{region}}\xspace}
\newcommand{\itregions}{\ensm{\mathit{regions}}\xspace}
\newcommand{\ttspecdecls}{\ensm{\itspec\text{-}\itdecls}\xspace}
\newcommand{\ttspecdecl}{\ensm{\itspec\text{-}\itdecl}\xspace}
\newcommand{\ttCallable}{\ensm{\mathit{Callable}}\xspace}
\newcommand{\ttcallable}{\ensm{\mathit{callable}}\xspace}
\newcommand{\ttFrame}{\ensm{\mathit{Frame}}\xspace}
\newcommand{\ttframe}{\ensm{\mathit{frame}}\xspace}
\newcommand{\ttframes}{\ensm{\mathit{frames}}\xspace}
\newcommand{\ttframec}{\ensm{\mathit{frame}_\mathit{map}}\xspace}
\newcommand{\ttframea}{\ensm{\mathit{frame}_\mathit{ale}}\xspace}
\newcommand{\ttmayframe}{\ensm{\mathit{may\text{-}frame}}\xspace}
\newcommand{\ttmustframe}{\ensm{\mathit{must\text{-}frame}}\xspace}
\newcommand{\ttpre}{\ensm{\mathit{pre}}\xspace}
\newcommand{\ttpost}{\ensm{\mathit{post}}\xspace}
\newcommand{\itmodules}{\ensm{\mathit{modules}}\xspace}
\newcommand{\itmodule}{\ensm{\mathit{module}}\xspace}
\newcommand{\itmap}{\ensm{\mathit{mapping}}\xspace}
\newcommand{\ttaptr}{\ensm{\mathit{Pointer}}\xspace}
\newcommand{\ttatype}{\ensm{\mathtt{type}}\xspace}
\newcommand{\ttafunc}{\ensm{\mathtt{func}}\xspace}
\newcommand{\ttaregion}{\ensm{\mathtt{region}}\xspace}
\newcommand{\ttaval}{\ensm{\mathtt{value}}\xspace}
\newcommand{\ttamaps}{\ensm{\mathit{Type Maps}}\xspace}
\newcommand{\ttadefs}{\ensm{\mathit{definition}}\xspace}
\newcommand{\ttaderive}{\ensm{\mathit{derivation}}\xspace}
\newcommand{\ttareq}{\ensm{\mathtt{require}}\xspace}
\newcommand{\ttaprov}{\ensm{\mathtt{provide}}\xspace}
\newcommand{\ttimport}{\ensm{\mathtt{import}}\xspace}
\newcommand{\ttalower}{\ensm{\mathtt{lower\text{-}with}}\xspace}
\newcommand{\ttfreads}{\ensm{\mathtt{read}}\xspace}
\newcommand{\ttfwrites}{\ensm{\mathtt{reg\text{-}modify}}\xspace}
\newcommand{\ttfmwrites}{\ensm{\mathtt{mem\text{-}modify}}\xspace}
\newcommand{\ttmodule}{\ensm{\mathtt{module}}\xspace}
\newcommand{\ttmayfreads}{\ensm{\mathtt{may\text{-}access}}\xspace}
\newcommand{\ttmayfwrites}{\ensm{\mathtt{may\text{-}modify}}\xspace}
\newcommand{\ttmustfreads}{\ensm{\mathtt{must\text{-}access}}\xspace}
\newcommand{\ttmustfwrites}{\ensm{\mathtt{must\text{-}modify}}\xspace}
\newcommand{\ttcasp}{\ensm{\mathit{casp}}\xspace}
\newcommand{\ttnil}{\ensm{\mathtt{nil}}\xspace}
\newcommand{\ttaloc}{\ensm{\mathtt{loc}}\xspace}
\newcommand{\ttnilptr}{\ensm{\mathit{Null}}\xspace}
\newcommand{\ttamap}{\ensm{\mathtt{Map}}\xspace}
\newcommand{\ttadt}{\ensm{\mathit{adt}}\xspace}
\newcommand{\ttmatch}{\ensm{\mathtt{match}}\xspace}
\newcommand{\ttwith}{\ensm{\mathtt{with}}\xspace}
\newcommand{\ttctor}{\ensm{\mathit{Ctor}}\xspace}
\newcommand{\ttisctor}{\ensm{\mathit{is}\text{-}\ttctor}\xspace}
\newcommand{\ttacases}{\ensm{\mathit{ale}\text{-}\mathit{match}\text{-}\mathit{cases}}}
\newcommand{\ttmatchor}{\ensm{\mathtt{\mid}}}
\newcommand{\ttany}{\ensm{\mathtt}{\_}}
\newcommand{\ttmaybe}{\ensm{\mathtt{Maybe}}}

\newcommand{\ttforall}{\ensm{\mathtt{forall}}\xspace}
\newcommand{\ttexists}{\ensm{\mathtt{exists}}\xspace}
\newcommand{\ttderef}{\ensm{\mathtt{deref\text{-}loc}}\xspace}

\newcommand{\ttsizeof}{\ensm{\mathtt{sizeof}}\xspace}

\newcommand{\ttmapemp}{\ensm{\ttnil}}

\newcommand{\bigtuple}{\ensm{\Pi}}
\newcommand{\bigsum}{\ensm{\Sigma}}

\newcommand{\aloctyp}{\ensm{\ttaloc}}
\newcommand{\amaptyp}[2]{\ensm{#1\tsp#2\tsp\ttamap}}

\newcommand{\alog}{\ensm{\phi}}
\newcommand{\auqonlylog}{\ensm{\alog_{\forall\:^*}}}
\newcommand{\auqlog}{\ensm{\alog_{\forall\:^*\exists\:^*}}}
\newcommand{\aeqlog}{\ensm{\alog_{\exists\:^*}}}
\newcommand{\aor}{\ensm{\vee}}
\newcommand{\aand}{\ensm{\wedge}}
\newcommand{\aimplies}{\ensm{\rightarrow}}
\newcommand{\aeq}{\ensm{=}}

\newcommand{\adtyp}{\ensm{\atyp_{\ttadt}}}
\newcommand{\casptyp}{\ensm{\atyp_{\ttcasp}}}
\newcommand{\loctyp}{\ensm{\atyp_{\ttloc}}}

\newcommand{\abitctor}{\ensm{\mathtt{Bitv}}\xspace}
\newcommand{\acptrctor}{\ttcPtr}
\newcommand{\abitvectortyp}{\ensm{\mathtt{bitvector}}}
\newcommand{\eMatch}[2]{\ensm{\ttmatch\tsp#1\tsp\ttwith\tsp#2}}
\newcommand{\ector}[1]{\ensm{\ttctor\tsp#1}}
\newcommand{\equant}{}
\newcommand{\eForall}[3]{\ensm{\ttforall\tsp #1\in #2. #3}}
\newcommand{\eExists}[3]{\ensm{\ttexists\tsp #1\in #2. #3}}

\newcommand{\tLet}[3]{\ensm{\ttlet\tsp\oftyp{#1}{#2}=#3}}
\newcommand{\tDef}[3]{\ensm{\ttdef\tsp{#1}\tsp{#2}=#3}}
\newcommand{\tDefs}[3]{\ensm{\ttdefs\tsp{#1}\tsp{#2}=#3}}

\renewcommand{\aprimitivetyp}{\atyp_{\mathit{base}}}
\newcommand{\afunctiontyp}{\atyp_{\mathit{func}}}
\newcommand{\aeffectfunctyp}{\atyp_{\mathit{proc}}}
\newcommand{\aregsettyp}{\atyp_{\mathit{regs}}}
\newcommand{\amemorytyp}{\atyp_{\mathit{mem}}}
\newcommand{\afuntyp}{\funtyp{\sseq{\cbasetypi}}{\aprimitivetyp}}
\newcommand{\abvtyp}{\bvtyp{\!}}
\newcommand{\awvectyp}[1]{\ensm{#1\ \ttwvec}}
\newcommand{\awptrtyp}[1]{\ensm{#1\ \ttwptr}}
\newcommand{\abvstyp}{\bvstyp{\!}}
\newcommand{\abvltyp}{\bvltyp{\!}}
\newcommand{\ttaset}{\abvstyp}
\newcommand{\aoption}{\ttC\ \vert\ \ttx}

\newcommand{\avectyp}{\mathtt{mem}}

\newcommand{\aderef}[1]{\ttderef(#1)}


\newcommand{\atenv}{\tenv}
\newcommand{\avenv}{\venv}
\newcommand{\agenv}{\genv}
\newcommand{\arholds}{\ensm{\vdash_a}\xspace}
\newcommand{\arenv}{\atenv,\agenv,\avenv}

\newcommand{\aenvrewrite}[4]{\ensm{#1\arholds#2\goesto_a #3\tholds #4}}
\newcommand{\arewrite}[2]{\ensm{\aenvrewrite{\arenv}{#1}{\tenv,\genv,\venv}{#2}}}

\newcommand{\kenv}{\ensm{\Sigma}\xspace}
\newcommand{\Kdeclxx}[3]{\ensm{#1 \tholds #2 \ttthen #3}}
\newcommand{\KdeclAA}[1]{\Kdeclxx{\kenv}{#1}{\kenv}}
\newcommand{\KdeclAB}[1]{\Kdeclxx{\kenv}{#1}{\kenv'}}
\newcommand{\KdeclBC}[1]{\Kdeclxx{\kenv'}{#1}{\kenv''}}
\newcommand{\KdeclAC}[1]{\Kdeclxx{\kenv}{#1}{\kenv''}}
\newcommand{\Kmachinex}[2]{\ensm{\tholds #1 \ttthen #2}}
\newcommand{\KmachineZ}[1]{\Kmachinex{#1}{\kenv_0}}
\newcommand{\KmachineA}[1]{\Kmachinex{#1}{\kenv}}
\newcommand{\KmachineB}[1]{\Kmachinex{#1}{\kenv'}}

\section{Introduction}
The goal of the \aquarium project is to synthesize the machine-dependent
parts of an operating system.
This has the potential to greatly reduce the amount of work needed
to port an operating system to a new machine architecture.
It also potentially reduces the depth of knowledge needed to do that
work -- currently an OS port requires deep expert knowledge of both
the OS and the new machine architecture -- and with luck the overall
amount of time involved as well.

This document does not discuss either code synthesis or OS porting in
any detail.
It is intended as a supplement to other project publications, which
should in general be read first.
While this document includes discussions of and rationales for
language features (as well as absence of language features) these
discussions assume familiarity with the surrounding context.

The two languages described in this document are \casp\footnote{
\casp is named for a jelly(fish) that features symbiotic
photosynthetic algae, which is for some reason spelled without 
the customary 'i'.
}, which is a
register-transfer-list-style machine description language used for writing
down the semantics of processor instruction sets, and \ale, which
is a specification or modeling language used for writing down
machine-independent specifications for assembly language code blocks
to be synthesized.

\casp also includes material for writing machine-dependent specifications
(this is the target for the \ale translator), a concept of mapping
modules used by the \ale translator, and a simple representation of
basic blocks in assembly language.
These are described in this document.

This document does not, however, describe any of the other languages
or file formats used in the \aquarium project or any of the \aquarium
tools.

The document is structured with six parts:
\begin{itemize}
\item The \casp language and abstract syntax (Sec. \ref{sec:lang})
\item \casp types (Section \ref{sec:check})
\item \casp semantics (Section \ref{sec:exec})
\item The \ale language and abstract syntax (Section \ref{sec:ale_lang})
\item \ale types and semantics (Section \ref{sec:ale_check})
\item \ale translation to \casp (Section \ref{sec:ale_exec})
\end{itemize}

\section{Notation}
In the abstract syntax, type judgments, and semantics judgments we use
italics for metavariables (e.g., $v$) and for words corresponding
to types in the abstract syntax (e.g., $\mathit{declaration}$).
We use \texttt{typewriter} font for words that correspond to language
keywords.
The notation $\overline{\alpha_i}$ means ``a sequence of one or more
$\alpha$, each to be referred to elsewhere as $\alpha_i$''.
If there are no references outside the overbar, the $i$ subscript may
be left off.
Epsilon ($\epsilon$) appearing in syntax represents an empty
production.
The notation \texttt{"}$\ldots$\texttt{"} represents a string literal
with arbitrary contents.

Bitvectors (machine integers) may be any width greater than zero.
Bitvector \emph{constants} are represented as \bitconstprefix{\ttC},
which can be thought of as an explicit sequence of zeros and
ones.
The number of bits in a bitvector constant (that is, the number of
digits) gives its type.
Thus, \bitconstprefix{00} and \bitconstprefix{0000} are different.
In the concrete syntax, bitvector constants whose size is a multiple
of 4 can also be written in the form \hexconst.
These are desugared in the parser and not shown further in this
document.

The \casp and \ale syntax should be considered disjoint.
Some elements are the same in each, but these are specified
separately regardless.
They use the same metavariables as well, which should not be mixed;
any language construct in a judgment should be all \casp or all
\ale.
In a few places mixing is needed, in which case the translation
defined in Section \ref{sec:ale_exec} is applied to allow inserting
\ale fragments into \casp terms.
The \ale rules in Section \ref{sec:ale_check} do use the same
\emph{environments} as the \casp rules.
These should be construed as holding only \casp elements.
Further details can be found in Section \ref{sec:ale_check}.



\section{\casp{} Overview}
\label{sec:lang}
This section covers the abstract syntax for \casp.

\casp is a register-transfer-list style language:
it models instructions as non-Turing-complete procedures that update a
machine state.
Its executable semantics are covered in Section \ref{sec:exec}.

We model the machine from the assembly-language programmer's perspective.
In particular, we do not treat memory as a huge block of address space
but instead treat it in small chunks passed in from somewhere else.
We model both control registers and general-purpose
registers as well as other machine state such as whether interrupts are enabled.

Furthermore we must allow assembler labels, which have addresses,
but those addresses are not resolved until after programs are compiled
and linked and must be treated as abstract.

\begin{figure*}
\centering
\begin{minipage}{0.50\textwidth}
\begin{align*}
	& & \textbf{(\casp Types)} \\
	& \atyp \bnfeq & \cbasetyp \bnfor \cmemtyp \bnfor \cfuntyp \\
	& & \\
	& \cbasetyp \bnfeq & \ttunit \bnfor \ttint \bnfor \ttbool \bnfor \ttstring \bnfor
	\tttypeid \bnfor \bvaltyp{\ttC} \\
	& \bnfor & \cregtyp \bnfor \cregstyp \bnfor \clbltyp\\
	& \cregtyp \bnfeq & \bvltyp{\ttC} \\
	& \cregstyp \bnfeq & \bvstyp{\ttC} \\
	& \clbltyp \bnfeq & \labeltyp{\ttC} \\  
	& & \\
	& \cmemtyp \bnfeq & \memtyp{\ttC_1}{\ttC_2}{\ttC_3} \\
	& \cfuntyp \bnfeq & \funtyp{\sseq{\cbasetypi}}{\cbasetyp} \\
	& & \\
	& & \textbf{(\casp{} Operators)} \\
	& \itunop \bnfeq & - \bnfor \bitsub\ \bnfor \neg \bnfor \bitnot \\\ 
	& \itbinop \bnfeq & =\ \bnfor\ \neq\ \bnfor\ + \bnfor - \bnfor * \bnfor / \ \bnfor\ <\ \bnfor\ <=\ \bnfor\ >\ \bnfor\ >= \\
	& \bnfor & \&\& \bnfor || \bnfor \xor \\
	& \bnfor & >> \bnfor {>>}_S \bnfor << \bnfor \bitand \bnfor \bitor \bnfor \bitxor \\
	& \bnfor & \bitadd \bnfor \bitsub \bnfor \bitmul \bnfor \bitdiv \\
	& \bnfor & \bitlt \bnfor \bitle \bnfor \bitgt \bnfor \bitge \\
	& \bnfor & \bitslt \bnfor \bitsle \bnfor \bitsgt \bnfor \bitsge\\
	& \bnfor & \cup \bnfor \cap \bnfor \subseteq \bnfor \setminus
\end{align*}
\end{minipage}
\hfill\vline\hfill
\begin{minipage}{0.48\textwidth}	
\begin{align*}
	& & \textbf{(\casp Values)} \\
	& \tildev \bnfeq & \ttv \bnfor \ttassertfalse \\
	& \ttv \bnfeq & \tttrue \bnfor \ttfalse \bnfor \ttC \bnfor \bitconst
	\bnfor \text{\strlit{}} \\
	& \bnfor & \ttr \bnfor \ptrform{\ttmemid}{\ttC} \\
	& & \\
	& & \textbf{(\casp{} Expressions)} \\
	& \tte \bnfeq & \tildev \bnfor \ttx \\
	& \bnfor & \strof{\tte} \\
	& \bnfor & \fapp{\ttfuncid}{\sseq{\tte}} \\
	& \bnfor & \itunop\ \tte\ \\
	& \bnfor & \tte_1\ \itbinop\ \tte_2\\
	& \bnfor & \bitidx{\tte}{\ttC} \bnfor \bitfield{\tte}{\ttC_1}{\ttC_2} \\
	& \bnfor & \eLet{\ttx}{\cbasetyp}{\tte_1}{\tte_2} \\ 
	& \bnfor &\sITE{\tte_1}{\tte_2}{\tte_3} \\ 
	& \bnfor & \ptrform{\ttmemid}{\tte} \\
	& \bnfor & \regderef{\tte} \bnfor \memread{\tte}\\
	& \bnfor & \NEW{\texttt{\ebranchto{\ttextid}}} \\
	& \bnfor & \setlit \\
	& \bnfor & \sizeofset{\tte} \bnfor \setmemberof{\tte_1}{\tte_2}
\end{align*}
\end{minipage}
\caption{\casp types, values, operators, and expressions.\label{fig:tr_casp_types1}}
\end{figure*}

\mypara{Notation}
We use the following metavariables:

\begin{tabular}{llr}
\ttx, \tty, \ttz & Program variables (binders) \\
\ttr            & Registers (abstract) \\
\ttC            & Integer constants (written in decimal) \\
\bitconst       & Bitvector constants (written in binary) \\
\atyp           & Types \\
\ttv            & Values \\
\tte            & Expressions \\
\ttS            & Statements \\
\tti, \ttj      & Rule-level integers \\
\end{tabular}

(Other constructions are referred to with longer names.)

A number of constructions are lists written out in longhand (with a
null case and a cons case) -- these are written out in longhand so
that typing and semantic judgments can be applied explicitly to each
case, in order to, for example, propaate environment updates correctly.

\mypara{Identifiers and Variables}
Identifiers are divided syntactically into \NNEW{seven} categories:
\begin{itemize}
\item \ttmemid are identifiers bound to memory regions, which are
second-class.
\item \ttfuncid are identifiers bound to functions, which are second-class.
\item \ttprocid are identifiers bound to procedures, which are second-class.
\item \ttxop are identifiers bound to instructions (``operations''), which
are akin to procedures but distinguished from them.
\item \tttypeid are identifiers for type aliases, which are bound to base types 
in declarations.
\item \ttxmoduleid are the names of ``modules'', which are used to select
among many possible groups of lowering elements.
\item Other identifiers \ttx are used for other things, and should be
assumed to \emph{not} range over the above elements.
\end{itemize}
Note that all identifiers live in the same namespace, and rebinding or
shadowing them is not allowed.
All these identifiers can be thought of as variables, in the sense
that they are names that stand for other things.
All of them are immutable once defined, including the ordinary
variables \ttx that contain plain values.

\mypara{Types}
Types are divided syntactically into base types (integers, booleans,
strings, bitvectors, etc.) and others (memory regions and functions).
User functions may handle only base types.
Furthermore, memory regions and functions are
second-class for reasons discussed below and are excluded in various
places in the syntax and the typing rules.

We use index typing to capture the bit width of values.

\mypara{Registers}
Registers are represented in the specification with the metavariable
$r$, which stands for the underlying abstract identity of a register.
Declaring a register, e.g., with 
$\ttlet\ttstate\ \ttx\ \ttcol\ \bvltyp{\ttC}$
as shown in \autoref{fig:tr_casp_types2},
allocates a fresh register $r$ and binds the variable \ttx to it.
A subsequent declaration of the form
$\ttlet\ \tty\ \ttcol\ \bvltyp{\ttC} = \ttx$
creates another variable \tty that refers to the same underlying
register.
One might think of registers as numbered internally.
We use the form $\ttlet\ttstate\ \ttcontrol\ \ttx\ \ttcol\ \bvltyp{\ttC}$ 
to declare specific \textit{control} registers, which are treated
differently by the framing rules.
The additional keyword \texttt{dontgate} inhibits state gating for the
register; this should be used for flags registers and anything similar
that is implicitly used by all ordinary code.

Some registers
have associated with them a
text form, which is declared
separately and is the form an assembler expects to parse.
The \synth uses this to extract an asssembly program from aquarium's
internal representation.
It is referred to by attaching the suffix \strof{} to the/a register
variable.
As some registers are not
directly addressable by the assembler (e.g., they might be subfields of some
larger addressable unit or nonaddressable internal state), not all
registers have a text form.
This is not readily checked statically, at least not without
introducing specific additional machinery, so invalid \strof{}
references fail at assembly code extraction time.

The type of a register is \bvltyp{\ttC}, which is a register that holds
a \ttC-bit bitvector.
The bitvector value in question can be updated by assigning a new
value; this is a statement ($\slAssign{\tte_1}{\tte_2}$) and can only
happen in places where
statements are allowed.
The construction $\regderef{\tte}$ reads a register.

The reader will note that the semantics rules for machines and
declarations do not provide initial values for registers.
Instead, executions are defined in terms of some initial register
state (and also some memory state), which is required to have the
right registers to match the machine definition.
This allows reasoning about the execution of programs and program
fragments in terms of many or all possible initial states.
These issues are discussed further below.

\mypara{Memory}
A memory region has the type \memtyp{\ttC_1}{\ttC_2}{\ttC_3}, shown in \autoref{fig:tr_casp_types1}.
This refers to a memory region that has $\ttC_2$ cells, each of which
stores a bitvector of width $\ttC_1$.
This memory region is addressed with pointers of width $\ttC_3$.
Note that we assume byte-addressed machines, and for the purposes of
both this specification and our implementation, we assume bytes are 8
bits wide.
(This restriction could be relaxed if we wanted to model various
historic machines.)
Thus a memory region of type \memtyp{32}{4}{32} has 4 32-bit values in
it, which can be addressed at byte offsets 0, 4, 8, and 12.
These values can of course be changed, like values in registers.

Memory regions are named with identifiers.
These names, and memory regions themselves, are not first class;
variables are not allowed to range over them.
Also note that memory regions are a property of programs (and thus are
declared in specifications) and not a property of the machine as a
whole.

\mypara{Pointers}
A pointer literal has the form \ptrform{\ttmemid}{\ttC}, in which
\ttmemid is the region name and \ttC is the offset, shown in \autoref{fig:tr_casp_types1}.
Because memory regions are second-class, \ttmemid must be specifically
one of the available declared memory regions.
Pointer literals exist in the abstract syntax, but are not allowed in
the concrete syntax except in specifications.
The only way to get a pointer value is to look up a label (discussed
immediately below) or have it provided in a register as part of the
initial machine state.

A pointer literal is treated as a bitvector of the same width, so one
can appear in a register or in memory.
However, we enforce a restriction (not captured in the semantics
rules so far) that no value in the initial machine state, whether in a
register or in memory, is a pointer unless required to be so by the
precondition part of the specification.
All other values are restricted to be plain bitvector values.

Addition and subtraction are allowed on pointers and they change the offset, 
but other bitvector operations (e.g., multiply) are disallowed
and fail.
Similarly, attempting to fetch from or store to a plain bitvector that
is not a pointer fails.
Note however that we do not statically distinguish pointers and plain
bitvectors.
(We could have used flow-sensitive typing to reason about when
registers and memory cells contain pointers and when they do not; but
this adds substantial complexity and for our problem domain does not
provide significant value.)
Instead, we step to failure at runtime.
This can be seen in the semantics rules.

Fetching from a pointer takes the form \memread{\tte}.
Storing to a pointer takes the form \smAssign{\tte_1, C}{\tte_2}.
The extra constant \ttC specifies the width of the cell pointed to.
(This is not an offset.)
Because we do not check pointers statically, we do not know the
memory region being pointed to and cannot look up its cell size; thus
we need the width explicitly for typing.
It is checked at runtime.

\begin{figure*}
\centering
\begin{minipage}{0.45\textwidth}
\begin{align*}
	& & \textbf{(\casp{} Statements)} \\
	& \ttS \bnfeq & \ttS\ttsemi\ \ttS \\
	& \bnfor & \papp{\ttprocid}{\sseq{\tte}} \\
	& \bnfor & \sLet{\ttx}{\cbasetyp}{\tte}{\ttS} \\
	& \bnfor & \sFor{\ttx}{\ttC_1}{\ttC_2}{\ttS} \\
	& \bnfor & \sITE{\tte}{\ttS_1}{\ttS_2} \\
	& \bnfor & \slAssign{\tte_1}{\tte_2} \\
	& \bnfor & \smAssign{\tte_1, C}{\tte_2} \\
	& \bnfor & \NEW{\sBRANCH{\tte}} \\
	& \bnfor & \ttassert(\tte) \\
	& \bnfor & \ttskip \\
	& \bnfor & \bferr
\end{align*}
\end{minipage}
\hfill\vline\hfill
\begin{minipage}{0.54\textwidth}
\begin{align*}
	& & \textbf{(\casp{} Declarations)} \\
	& \decls \bnfeq\ & \eps\ \bnfor \decl \ttsemi\ \decls \\
	& \decl \bnfeq & \tttype\ \tttypeid = \cbasetyp \\
	& \bnfor & \ttlet\ \ttx\ \ttcol\ \cbasetyp = \tte \\
	& \bnfor & \ttlet\ \strof{\ttx} = \text{\tte} \\
	& \bnfor & \ttdef\ \ttfuncid\ \funtyp{\sseq{\ttx_i\ \ttcol\ \cbasetypi}}{\cbasetyp} = \tte \\
	& \bnfor & \ttproc\ \ttprocid\ \proctyp{\sseq{\ttx_i\ \ttcol\ \cbasetypi}} = \ttS \\
	& \bnfor & \ttlet\ttstate\ \ttx\ \ttcol\ \cregtyp \\ 
	& \bnfor & \ttlet\ttstate\ \ttcontrol\ \ttx\ \ttcol\ \cregtyp \\ 
	& \bnfor & \ttlet\ttstate\ \ttcontrol\ \ttdontgate\ \ttx\ \ttcol\ \cregtyp \\ 
	& \bnfor & \ttlet\ttstate\ \ttmemid\ \ttcol\ \cmemtyp \\
	& \bnfor & \ttlet\ttstate\ \ttmemid\ \ttcol\ \cmemtyp\ \ttwith\ \NNEW{\ttx} \\
	& & \\
	& \defops \bnfeq\ & \eps \bnfor \defop \ttsemi\ \defops \\
	& \defop \bnfeq\ & \ttdefop\ \ttxop\  \ttinbrace{\tttxt=\tte\ttcomma\ttsem=\ttS} \\
	& \bnfor & \ttdefop\ \ttxop\ \sseq{\ttx_i\ \ttcol\ \cbasetypi}\ \ttinbrace{\tttxt=\tte\ttcomma\ttsem=\ttS}
\end{align*}
\end{minipage}
\caption{\casp statements and declarations.\label{fig:tr_casp_types2}}
\end{figure*}

\mypara{Labels}
As mentioned above, the term ``label'' means an assembler label or linker
symbol.
These have addresses (or depending on how one looks at them, they are
addresses) and those addresses are constants, but the constants are
not known at assembly time, so we must model them abstractly.

When one declares a memory region, one may attach a label to it, which
is an additional identifier.
This identifier is created as a variable of type \labeltyp{\ttC} in \autoref{fig:tr_casp_types1}.
The value is a pointer to the first entry in the region, and a single
type subsumption rule allows this value to be accessed and placed in
an ordinary register or variable of suitable bitvector type.
The intended mechanism is that for each machine the preferred
instruction on that machine for loading assembler symbols into a
register can be defined to take an operand of type \labeltyp{\ttC},
and then its value can be assigned to the destination register.
This type restriction on the operand is sufficient to synthesize
programs that use labels correctly.

\mypara{\NEW{Code Positions and Branching}}
\casp code blocks may contain branches, either forward within the
block (branching backwards is forbidden) or to a single external
assembler label outside the block.
This model is sufficient for our block-oriented synthesis approach;
more complex control flow can be handled with multiple blocks.

Consequently, a branch may either go to the external assembler label
(which terminates execution of the current block) or skip zero or more
instructions.
We model branch targets as an 8-bit skip count.
In \autoref{fig:tr_casp_types2}, The statement \sBRANCH{\bitconst{}} skips \bitconst{} instructions;
\sBRANCH{\texttt{0xff}} jumps to the external label.
This statement may be used to define both conditional and
unconditional branch instructions.
Such instructions should be defined to take an operand of type
\bvtyp{8} to choose the branch destination.
This magic number should then be printed to the output assembly text
using the built-in function \texttt{textlabel}, which replaces it with
a valid assembler label, either the selected external label string or
a scratch label attached to the proper destination instruction.

Specifications do not need to be directly concerned with internal
branches, which occur or not as needed.
However, external branches are part of a block's specification;
typically the purpose of a block with an external branch is to test
some condition and then either branch away or fall through to the next
block.
It is thus necessary to be able both to name the external label to
use and to specify the conditions when it should be reached.
For this purpose a predicate \texttt{branchto} is provided.
It may appear in the postcondition and governs the exit
path from the block: if forced to true, the block branches to the
external assembler label, and if false, the block falls through from
its last instruction.
The concrete syntax is
\texttt{branchto(dest)}
which also sets the assembler label used to \texttt{dest}.
It is not valid to name more than one such assembler label.

Note that the assembler labels used in branching are, though also
assembler labels, a separate mechanism unrelated to the labels
attached to memory regions; they are code labels rather than data
labels and inherently work differently.

Also note that the implementation is slightly more
complicated than the description given above and the rules given in
the semantics section.
In the implementation, the magic number that triggers an external
branch is not 255 but the number $l$ such that $l$ added to the
instruction number of the branch adds to 99.
This detail affects only intermediate synthesis results in the
internal format; the extracted assembly code is the same.

\mypara{Register Sets}
Register sets are second-class elements intended to exist only as
literals and only as the result of lowering machine-independent
specifications that cannot directly talk about specific registers.
Currently they do not exist in the implementation.
Register sets are not allowed to be operands to instructions to avoid state
explosions when synthesizing.
This restriction is currently not captured in the abstract syntax or
typing rules.

\mypara{Functions and Procedures}
Functions, defined with \ttdef in \autoref{fig:tr_casp_types2}, are pure functions whose bodies are
expressions.
They produce values.
They can access registers and memory, and can fail, but cannot update
anything.
Procedures, defined with \ttproc, are on the other hand
impure and their bodies are statements.
They do not produce values, but they may update the machine state.
They are otherwise similar, and are intended to be used to abstract
out common chunks of functionality shared among multiple instructions
in machine descriptions.
Functions can also be used for state hiding in specifications.

Functions and procedures are second-class; they may be called only by
their own name and may not be bound to variables or passed around.
Furthermore, they are only allowed to handle base types: higher-order
functions are explicitly not supported.

\mypara{Operations}
Operations (defined with \ttdefop in \autoref{fig:tr_casp_types2}) are essentially instructions, and we
refer to these
interchangeably.
An operation takes zero or more operands and transforms
the machine state as defined by one or more statements.
Operands are currently defined as expressions, but are restricted as
follows:
\begin{itemize}
\item They may be values, but not string values, and not \exprfail.
\item They may be variables of register type.
\item They may be variables of label type.
\end{itemize}
This restriction affects what the synthesizer tries to generate; a
broader set of expressions may be accepted for verification or
concrete execution and simply evaluated in place.

There is an important distinction between ``operations'' and
``instructions''.
Operations are the units in which \casp{} reasons about machine
operations and the units in which \casp{} generates programs and code
fragments, but they do not necessarily need to be single instructions.
The text output to the assembler is arbitrary and can be computed on
the fly based on the operand value.
On some platforms the assembler defines so-called ``synthetic
instructions'' that are potentially multiple real instructions.
This facility takes that a step further by allowing the writer of the
machine description to define their own synthetics.


\mypara{Other Constructs}
\bitidx{\tte}{\ttC} and \bitfield{\tte}{\ttC_1}{\ttC_2} in \autoref{fig:tr_casp_types1} extract a
single bit and a slice, respectively, from a bitvector.
The offsets are constants; a shift can be used beforehand if variable
offsets are needed.
The width of the slice must be constant for static typing.

\begin{figure*}
\centering
\begin{minipage}{0.49\textwidth}
\begin{align*}
	& & \textbf{(\casp{} Machines)} \\
	& \machine \bnfeq & \decls\ttsemi\ \defops \\
	& & \\
	& & \textbf{(\casp{} Programs)} \\
	& \itinst \bnfeq & \ttxop \bnfor \ttxop\ \sseq{\tte}\\
	& \itinsts \bnfeq & \eps \bnfor \itinst \ttsemi\ \itinsts \\
	& \program \bnfeq & \itinsts
\end{align*}
\end{minipage}
\hfill\vline\hfill
\begin{minipage}{0.49\textwidth}
\begin{align*}
	& & \textbf{(\casp{} Lowerings)} \\
	& \itmodules \bnfeq & \eps \bnfor \itmodule \ttsemi\ \itmodules \\
	& \itmodule \bnfeq & \ttmodule\ \ttxmoduleid\ \ttinbrace{\ttdecls \ttsemi\ \ttframe} \\
	& & \\
	& & \textbf{(\casp{} Specifications)} \\
	& \ttframe \bnfeq & \ttfwrites\ \ttcol\ \sseq{\ttregidi} \\
	& \bnfor & \ttfmwrites\ \ttcol\ \sseq{\ptrform{\ttmemidi}{\tte_i}}\\
	& \ttframes \bnfeq & \eps \bnfor \ttframe\ \ttframes \\
	& \ttpre \bnfeq & \tte \\ 
	& \ttpost \bnfeq & \tte \\
	& \itspec \bnfeq & \ttdecls \ttsemi\ \ttframes \ttsemi\ \ttpre \ttsemi\ \ttpost
\end{align*}
\end{minipage}
\caption{\casp machines, lowerings, programs, and specifications.\label{fig:tr_casp_types3}}
\end{figure*}


\mypara{Machines, Lowering, Specifications, and Programs}

A \machine~ is the full description of a machine architecture; it
includes declarations (including types, constants, registers,
functions and procedures) and also instructions, shown in \autoref{fig:tr_casp_types3}.
This is typically a standalone file but may be a set of files referenced via
\ttinclude.

A (single) \lowering is a collection of declarations used to instantiate
elements in \ale translations, shown in \autoref{fig:tr_casp_types3}.
These are placed into a \itmodule, with multiple modules per file, so
that the \lowerings associated with multiple related \ale
specifications can be kept together.
The \ttimport \ttxmoduleid directive enables sharing common elements in
one module \ttxmoduleid across multiple specifications.
The modules used to lower a specification are selected using the
\ttalower declaration in \ale.

A \itspec (specification) is a precondition and postcondition, which are boolean
expressions, along with optional permission to destroy additional
registers (the \ttframe), shown in \autoref{fig:tr_casp_types3}.
\casp specifications are produced by compiling \ale specifications.
Note that a module can also contain frame declarations; these are
added to any provided in the \ale specification.
A code block is permitted to destroy any register that is either
explicitly listed in the frame declarations or mentioned in the
postcondition, while it may read any register mentioned in the
precondition and any control register.
This restriction is currently not adequately captured in the
semantics rules.

A \program\ is a sequence of instruction invocations, shown in \autoref{fig:tr_casp_types3}.

\mypara{Built-in Functions}
Here is a partial list of the built-in functions in \casp{}.
\begin{itemize}
	\item \texttt{empty} (\funtyp{\ttint\ \ttC}{\bvstyp{\ttC}}) produces an empty register set of the requested bit size.
	\item \texttt{hex} (\funtyp{\bvtyp{\ttC}}{\ttstring}) prints numbers in hexadecimal.
	\item \texttt{bin} (\funtyp{\bvtyp{\ttC}}{\ttstring}) prints numbers in binary.
	\item \texttt{dec} (\funtyp{\bvtyp{\ttC}}{\ttstring}) prints numbers in decimal.
	\item \texttt{lbl} (\funtyp{\labeltyp{\ttC}}{\ttstring})
	prints labels (it returns the label identifier as a string).
	This is for data labels attached to memory locations.
	\item \NEW{\texttt{textlabel} (\funtyp{\bvtyp{8}}{\ttstring})}
	prints branch offsets as assembler labels.
	This is for code labels, as described above.
	\item \texttt{format} (\funtyp{\ttstring}{\funtyp{\ttstring\
	$\ldots$}{\ttstring}}) formats strings.
	The first argument is a format string; the remainder of the
	arguments are substituted into the format string where a
	dollar sign appears followed by the argument number (1-based).
	(A literal dollar sign can be inserted by using
	\texttt{\$\$}.)
	The number of additional arguments expected is deduced from
	the contents of the format string.
	\item \texttt{bv\_to\_len}
	(\funtyp{\ttint\ \ttC_1}{\funtyp{\bvtyp{\ttC_2}}{\bvtyp{\ttC_1}}})
	returns a new bitvector of size $\ttC_1$ with the same value
	up to the ability of the new size to represent that value.
	\item \texttt{bv\_to\_uint} (\funtyp{\bvtyp{\ttC_1}}{\ttint})
	converts a bitvector to unsigned int.
	\item \texttt{uint\_to\_bv\_l} (\funtyp{\ttint\
	\ttC_1}{\funtyp{\ttint\ \ttC_2}{\bvtyp{\ttC_1}}}) converts an
	unsigned int $\ttC_2$ into a bitvector of size $\ttC_1$.
	\item \texttt{isptr} (\funtyp{\bvtyp{\ttC}}{\ttbool}) tests
	at runtime if a bitvector value is a pointer or not.
\end{itemize}

Note that some of these functions have their own typing rules, some of
which are polymorphic in bitvector size.
We have not complicated the typing rules presented by including all of
these as special cases.

\mypara{Concrete Syntax}
We do not describe the concrete syntax in detail here; however,
it does not stray very far from the abstract syntax.
The operator precedence and most of the operator spellings are taken
from C but most of the rest of the concrete syntax is ML-style.
There are also a few things desugared in the parser and not shown in the
abstract syntax.
As already mentioned, bitvector constants whose size is a multiple of 4 can
also be written in the form \hexconst.
Syntax of the form \hexof{e}, \binof{e}, and \decof{e} is
converted to the built-in functions \tthex, \ttbin, \ttdec
respectively.
These print either integers or bitvectors as strings in hexadecimal,
binary, or decimal respectively.
The syntax \lblof{\NNEW{\ttx}} is similarly converted to the built-in
function \ttlbl.
This produces the label (that is, the identifier naming the label) as
a string.
Further the concrete syntax supports include files via an \ttinclude
directive, which is useful for sharing common elements.

\section{\casp{} Static Typing}
\label{sec:check}

\begin{figure*}
\textbf{(Type Well-Formedness)}\\
\begin{tabular}{cccc}
\irule{~}{\wfenv\twfholds\ttunit} 	 &  \irule{~}{\wfenv\twfholds\ttint} &
\irule{~}{\wfenv\twfholds\ttbool}	&  \irule{~}{\wfenv\twfholds\ttstring}
\end{tabular}
\nextrulesmall
\begin{tabular}{cccc}
\irule{\ttC > 0}
	{\wfenv\twfholds\bvaltyp{\ttC}}
&  
\irule {\ttC > 0}
	{\wfenv\twfholds\bvltyp{\ttC}}
&
\irule {\ttC > 0}
	{\wfenv\twfholds\labeltyp{\ttC}}
& 
\irule{\ttC > 0}
	{\wfenv\twfholds\bvstyp{\ttC}}
\end{tabular}
\nextrulesmall
\begin{tabular}{ccc}
\irule
	{\ttC_1 > 0,\ \ttC_2 > 0,\ \ttC_3> 0}
	{\wfenv\twfholds\memtyp{\ttC_1}{\ttC_2}{\ttC_3}}
&
\irule{\genv(\ttx)=\atyp\nextclause\\ 
	\wfenv\twfholds\atyp}
	{\wfenv\twfholds\ttx}
& 
\irule
	{\forall i, \wfenv\twfholds\atyp_i \nextclause\\ 
	\wfenv \twfholds \atyp_r}
	{\wfenv\twfholds\funtyp{\sseq{\atyp_i}}{\atyp_r}}
\end{tabular}
\caption{\casp type well-formedness.\label{fig:tr_casp_type_wf}}
\end{figure*}




This section describes the \casp type system.

\mypara{Environments}
The type system uses two environments: $\genv$ maps type alias names
to the types they represent, and $\tenv$ maps variables to the types
assigned to them.
Recall from the syntax that only base types may have alias names, so
alias names can be treated as base types.

\mypara{Well-Formedness}
Since types include alias names, we need to check that a proposed
alias name is actually a type name.
At the same time we insist that the widths of bitvectors be greater
than zero.
The judgment for this has the form $\wfenv \twfholds \atyp$, shown in \autoref{fig:tr_casp_type_wf}.
There is an intended invariant that only well-formed types may be
entered into the variable typing environment $\tenv$, so that types
taken out of it do not need to be checked for well-formedness again.

In a typing environment comprised of $\genv$ mapping user-defined type
names (type aliases) to types and $\tenv$ mapping program binders
(variables) to types, we say that a type is well formed when all type
names are well-formed and all indices are of type $\ttint$ and nonzero.

\begin{figure*}
\textbf{(Expression Typing)}\\
\begin{tabular}{cccc}
\irule{~}
		{\typenv\tholds\hastyp{\ttC}{\ttint}}
&
\irule{~}
		{\typenv\tholds\hastyp{\text{\strlit{}}}{\ttstring}}
&
\irule{~}
		{\typenv\tholds\hastyp{\tttrue}{\ttbool}}
&
\irule{~}
		{\typenv\tholds\hastyp{\ttfalse}{\ttbool}}
\end{tabular}
\nextrulesmall
\begin{tabular}{cccc}
\irule
	{\ttC = {\{0,1\}}^\ttk}
	{\typenv\tholds\hastyp{\bitconst}{\bvtyp{\ttk}}}
&
\irule
	{\wfenv\twfholds\atyp}
	{\typenv \tholds\hastyp{\ttassertfalse}{\atyp}}
&
\irule
	{\tenv(\ttx) = \atyp\nextclause \\ 
	\wfenv\twfholds\atyp}
	{\typenv\tholds\hastyp{\ttx}{\atyp}}
&
\irule
{ \typenv \tholds \tte\ \ttcol\ \cregtyp}
{\typenv\tholds\hastyp{\strof{\tte}}{\ttstring}}
\end{tabular}
\nextrulesmall
\begin{tabular}{cc}
\irule
	{\typenv\tholds\hastyp{\ttfuncid}{\left( \funtyp{\sseq{\cbasetypi}}{\cbasetyp} \right)}\nextclause \\
		\forall \tti,\ \typenv\tholds\hastyp{\tte_i}{\cbasetypi}}
	{\typenv\tholds\hastyp{\fapp{\ttfuncid}{\sseq{\tte_i}}}{\cbasetyp}}
&
\irule
	{\typenv\tholds\hastyp{\tte}{\cbasetypone} \nextclause \\
		\dholds \hastyp{\itunop}{\funtyp{\cbasetypone}{\cbasetyptwo}} }
	{\typenv\tholds\hastyp{\itunop\ \tte}{\cbasetyptwo}}
\end{tabular}

\nextrulesmall
\irule
	{\typenv\tholds\hastyp{\tte_1}{\cbasetypone}\nextclause \\ 
		\typenv\tholds\hastyp{\tte_2}{\cbasetypone} \nextclause \\
		\dholds \hastyp{\itbinop}{\funtyp{\funtyp{\cbasetypone}{\cbasetypone}}{\cbasetyptwo}} }
	{\typenv\tholds\hastyp{\tte_1\ \itbinop\ \tte_2}{\cbasetyptwo}}

\nextrulesmall
\begin{tabular}{cc}
\irule
	{\typenv\tholds\hastyp{\tte}{\bvtyp{\ttC_2}}\nextclause \\ 
		0 \le \ttC_1 < \ttC_2 }
	{\typenv\tholds\hastyp{\bitidx{\tte}{\ttC_1}}{\bvtyp{1}}}
&
\irule
	{\typenv\tholds\hastyp{\tte}{\bvtyp{\ttC_3}}\nextclause \\ 
		0 \le \ttC_1 < \ttC_2 \le \ttC_3 \nextclause \\
		\ttk = \ttC_2-\ttC_1}
	{\typenv\tholds\hastyp{\bitfield{\tte}{\ttC_1}{\ttC_2}}{\bvtyp{\ttk}}}
\end{tabular}

\nextrulesmall
\irule
	{\typenv\tholds\hastyp{\tte_1}{\cbasetyp}\nextclause \\ 
		\ttx\notin\typenv\nextclause \\ 
		\genv, \tenv[\ttx\envgoesto\cbasetyp]\tholds\hastyp{\tte_2}{\atyp_2} }
	{\typenv\tholds\hastyp{\eLet{\ttx}{\cbasetyp}{\tte_1}{\tte_2}}{\atyp_2}}
	
\nextrulesmall
\irule
	{\typenv\tholds\hastyp{\ttb}{\ttbool}\nextclause \\ 
		\typenv\tholds\hastyp{\tte_{1}}{\atyp} \nextclause \\ 
		\typenv\tholds\hastyp{\tte_{2}}{\atyp}}
	{\typenv\tholds\hastyp{\sITE{\ttb}{\tte_1}{\tte_2}}{\atyp}}

\nextrulesmall
\begin{tabular}{cc}
\irule
{\typenv\tholds\hastyp{\tte}{\ttint}\nextclause \\ 
	\typenv\tholds\hastyp{\ttmemid}{\memtyp{\_}{\_}{\ttC}}}
{\typenv\tholds\hastyp{\ptrform{\ttmemid}{\tte}}{\cptrtyp{\ttC}}}
&
\irule
	{\typenv\tholds\hastyp{\tte}{\bvaltyp{\_}}\nextclause \\ \ttC > 0}
	{\typenv\tholds\hastyp{\fetch{\tte}{\ttC}}{\bvaltyp{\ttC}}}
\end{tabular}

\nextrulesmall
\begin{tabular}{ccc}
\irule
	{\typenv\tholds\hastyp{\tte}{\bvltyp{\ttC}}}
	{\typenv\tholds\hastyp{\regderef{\tte}}{\bvaltyp{\ttC}}}
&
\irule
	{\typenv\tholds\hastyp{\NNEW{\ttx}}{\labeltyp{\ttC}}}
	{\typenv\tholds\hastyp{\NNEW{\ttx}}{\bvaltyp{\ttC}}}
&
\irule{~}
	{\NEW{\typenv\tholds\hastyp{\ebranchto{\ttextid}}{\ttbool}}}
\end{tabular}

\nextrulesmall
\begin{tabular}{ccc}
\irule
	{\forall \tti \in(1\ldots\ttk),\ \typenv\tholds \hastyp{\ttx_i}{\bvltyp{\ttC}}}
	{\typenv\tholds\hastyp{\ttinbrace{\ttx_1, \ldots, \ttx_k}}{\bvstyp{\ttC}}}
&
\irule
	{\typenv\tholds\hastyp{\tte}{\bvstyp{\ttC}}}
	{\typenv\tholds\hastyp{\sizeofset{\tte}}{\ttint}}
&
\irule
	{\typenv\tholds\hastyp{\tte_1}{\bvltyp{\ttC}}\ \ \ 
		\typenv\tholds\hastyp{\tte_2}{\bvstyp{\ttC}}}
	{\typenv\tholds\hastyp{\setmemberof{\tte_1}{\tte_2}}{\ttbool}}
\end{tabular}
\caption{\casp typing rules for expressions.\label{fig:tr_casp_type_expr}}
\end{figure*}

\mypara{Expressions}
Expressions produce values that have types.
Because types appear explicitly in some expressions (e.g., \texttt{let}),
we need both environments, so the form of an
expression typing judgment is
$\typenv\tholds\hastyp{\tte}{\atyp}$, shown in \autoref{fig:tr_casp_type_expr}.
This means that we conclude \tte has type \atyp.
Note that the \strof{} form is restricted to registers; it is
specifically for extracting the assembly text form of a register.
We have not written out a separate rule for each unary and binary
operator.
The types of operators are as shown in \autoref{fig:tr_caps_ops}.
Note that the bitvector operators are polymorphic in bit size.

\begin{figure*}
\newcommand{\funtypp}[3]{\funtyp{\funtyp{#1}{#2}}{#3}}
\newcommand{\bvC}{\bvaltyp{\ttC}}
\newcommand{\bvsC}{\bvstyp{\ttC}}
\newcommand{\allC}{$\forall \ttC,$}
\begin{tabular}{ccccr}
$-$      &&&		& \funtyp{\ttint}{\ttint} \\
\bitsub	 &&&		& \allC\funtyp{\bvC}{\bvC} \\
$\neg$	 &&&		& \funtyp{\ttbool}{\ttbool} \\
\bitnot	 &&&		& \allC\funtyp{\bvC}{\bvC} \\
~	 &&&		& ~ \\
$=$ & $\neq$	&&	& $\forall \cbasetyp,$ \funtypp{\cbasetyp}{\cbasetyp}{\ttbool} \\
$+$ & $-$ & $*$ & $/$		& \funtypp{\ttint}{\ttint}{\ttint} \\
$<$ & $<=$ & $>$ & $>=$		& \funtypp{\ttint}{\ttint}{\ttbool} \\
$\&\&$ & $||$ & $\xor$	&& \funtypp{\ttbool}{\ttbool}{\ttbool} \\
$>>$ & ${>>}_S$ & $<<$	&& \allC\funtypp{\bvC}{\bvC}{\bvC} \\
\bitand&\bitor&\bitxor	&& \allC\funtypp{\bvC}{\bvC}{\bvC} \\
\bitadd&\bitsub&\bitmul&\bitdiv & \allC\funtypp{\bvC}{\bvC}{\bvC} \\
\bitlt&\bitle&\bitgt&\bitge & \allC\funtypp{\bvC}{\bvC}{\ttbool}\\
\bitslt&\bitsle&\bitsgt&\bitsge &\allC\funtypp{\bvC}{\bvC}{\ttbool}\\
$\cup$&$\cap$&$\setminus$	&& \allC\funtypp{\bvsC}{\bvsC}{\bvsC}\\
$\subseteq$			&&&& \allC\funtypp{\bvsC}{\bvsC}{\ttbool}
\end{tabular}
\caption{\casp typing rules for \itunop and \itbinop.}\label{fig:tr_caps_ops}
\end{figure*}

Arguably the right hand argument of the shift operators should be
allowed to be a different width.
There is one rule for pointer literals that covers both the expression
and the value form.
There is no rule (either in the typing or in the semantics) that
allows taking a subrange of a memory region as a new smaller region.
We have not needed this for our use cases, and keeping the set of
possible regions fixed simplifies a number of things.

\begin{figure*}
\textbf{(Statement Typing)}
\nextrulesmall
\begin{tabular}{cc}
\irule
	{\typenv\tholds{\ttS_1}\nextclause \\ 
		\typenv\tholds{\ttS_2} }
	{\typenv\tholds{\ttS_1;\ \ttS_2}}
&
\irule
	{\typenv\tholds\hastyp{\ttprocid}{\left( \proctyp{\sseq{\cbasetypi}} \right)}\nextclause \\
		\forall \tti,\ \typenv\tholds\hastyp{\tte_i}{\cbasetypi}}
	{\typenv\tholds\papp{\ttprocid}{\sseq{\tte_i}}}
\end{tabular}

\nextrulesmall
\begin{tabular}{cc}
\irule
	{\typenv\tholds\hastyp{\tte}{\cbasetyp}\nextclause \\ 
		\ttx\notin\typenv\nextclause \\  
		\typenv[\ttx\envgoesto\cbasetyp]\tholds{\ttS} }
	{\typenv\tholds{\sLet{\ttx}{\cbasetyp}{\tte}{\ttS}}}
&
\irule
	{\ttx\notin\typenv\nextclause \\
		\typenv[\ttx\envgoesto\ttint]\tholds{\ttS}\nextclause}
	{\typenv\tholds{\sFor{\ttx}{\ttC_1}{\ttC_2}{\ttS}}}
\end{tabular}

\nextrulesmall
\begin{tabular}{cc}
\irule
	{\typenv\tholds\hastyp{\tte}{\ttbool}\nextclause \\ 
		\typenv\tholds{\ttS_{1}}\nextclause \\ 
		\typenv\tholds{\ttS_{2}}}
	{\typenv\tholds{\ \sITE{\tte}{\ttS_1}{\ttS_2}}}
&
\irule
	{\typenv\tholds\hastyp{\tte_1}{\bvltyp{\ttC}}\nextclause \\ 
		\typenv\tholds\hastyp{\tte_2}{\bvaltyp{\ttC}}}
	{\typenv\tholds\slAssign{\tte_1}{\tte_2}}
\end{tabular}

\nextrulesmall
\begin{tabular}{ccc}
\irule
	{\typenv\tholds\hastyp{\tte_1}{\bvaltyp{\ttC_1}}\nextclause \\ 
		\typenv\tholds\hastyp{\tte_2}{\bvaltyp{\ttC_2}}}
	{\typenv\tholds\smAssign{\tte_1, \ttC_2}{\tte_2}}
&
\NEW{
	\irule
	{\typenv\tholds\hastyp{\tte}{\bvtyp{8}}}
	{\typenv\tholds{\sBRANCH{\tte}}}
}
&
\irule
	{\typenv\tholds\hastyp{\tte}{\ttbool}}
	{\typenv\tholds{\ttassert(\tte)}}
\end{tabular}
	
\begin{tabular}{cc}
\irule{~}
		{\typenv\tholds{\ttskip}}
&
\irule{~} 
		{\typenv\tholds{\bferr}}
\end{tabular}
\caption{\casp typing rules for statements.\label{fig:tr_casp_type_stmt}}
\end{figure*}

\mypara{Statements}
Statements do not produce values.
We still need both environments, though, so the form of a typing
judgment for a statement is $\typenv\tholds{\ttS}$, shown in \autoref{fig:tr_casp_type_stmt}.
This means that \ttS is well typed.

\begin{figure*}
\textbf{(Declaration Typing)}

\nextrulesmall
\begin{tabular}{cc}
\irule {~}
		{\genv, \tenv\tholds \eps \ttthen \genv, \tenv}
&
\irule 
	{\genv, \tenv\tholds \decl \ttthen \genv', \tenv' \nextclause \\ 
		\genv', \tenv' \tholds \decls \ttthen \genv'', \tenv''}
	{\genv,\tenv\tholds \decl \ttsemi\ \decls \ttthen \genv'',\tenv''}
\end{tabular}

\nextrulesmall
\begin{tabular}{cc}
\irule
	{\wfenv\twfholds\cbasetyp\nextclause \\ 
		\tttypeid\notin\genv, \env\nextclause \\
		\genv'= \genv[\tttypeid\goesto\cbasetyp]}
	{\genv, \tenv\tholds \tttype\ \tttypeid = \cbasetyp \ttthen \genv',\tenv }
&
\irule
	{\typenv \tholds \ttx\ \ttcol\ \cregtyp\nextclause \\
		\genv,\tenv\tholds\hastyp{\tte}{\ttstring} }
	{\genv,\tenv\holds\ttlet\ \strof{\ttx} = \tte\ttthen\genv, \tenv}
\end{tabular}
\nextrulesmall
\irule
	{\wfenv\twfholds\cbasetyp\nextclause \\ 
		\ttx\notin\genv,\tenv\nextclause \\
		\genv, \env\tholds\hastyp{\tte}{\cbasetyp}\nextclause \\ 
		\tenv'=\tenv[\ttx\envgoesto\cbasetyp]}
	{\genv,\tenv\holds\ttlet\ \ttx\ \ttcol\ \cbasetyp = \tte\ttthen\genv, \tenv'}

\nextrulesmall
\irule
	{\wfenv\twfholds \left(\funtyp{\sseq{\cbasetypi}}{\cbasetyp}  \right)\nextclause \\
		\ttfuncid\notin\genv,\tenv \nextclause \\
		\tenv' = \tenv[\forall \tti, \ttx_i\envgoesto\cbasetypi]\nextclause \\
		\genv, \env'\tholds\hastyp{\tte}{\cbasetyp} \nextclause \\
		\tenv'' = \tenv[\ttfuncid\envgoesto\left( \funtyp{\sseq{\ttx_i\ \ttcol\ \cbasetypi}}{\cbasetyp} \right)]}
	{\genv,\tenv\holds\ttdef\ \ttfuncid\ \funtyp{\sseq{\ttx_i\ \ttcol\ \cbasetypi}}{\cbasetyp} = \tte\ttthen\genv, \tenv''}

\nextrulesmall
\irule
{\wfenv\twfholds \left( \proctyp{\sseq{\cbasetypi}} \right)\nextclause \\
	\ttprocid\notin\genv,\tenv \nextclause \\
	\tenv' = \tenv[\forall \tti, \ttx_i\envgoesto\cbasetypi]\nextclause \\
	\genv, \env'\tholds\ttS \nextclause \\
	\tenv'' = \tenv[\ttprocid\envgoesto \left( \proctyp{\sseq{\ttx_i\ \ttcol\ \cbasetypi}} \right)]}
{\genv,\tenv\holds\ttproc\ \ttprocid\ \proctyp{\sseq{\ttx_i\ \ttcol\ \cbasetypi}} = \ttS\ttthen\genv, \tenv''}

\nextrulesmall
\irule
	{\wfenv\twfholds\bvltyp{\ttC}\nextclause \\ 
		\ttx\notin\genv,\tenv \nextclause \\
		\tenv' = \tenv[\ttx\envgoesto\bvltyp{\ttC}]}
	{\genv, \tenv\holds\ttlet\ttstate\ \ttx\ \ttcol\ \bvltyp{\ttC}\ttthen \genv, \tenv'}
	
\nextrulesmall
\irule
{\wfenv\twfholds\bvltyp{\ttC}\nextclause \\ 
	\ttx\notin\genv,\tenv \nextclause \\
	\tenv' = \tenv[\ttx\envgoesto\bvltyp{\ttC}]}
{\genv, \tenv\holds\ttlet\ttstate\ \ttcontrol\ \ttx\ \ttcol\ \bvltyp{\ttC}\ttthen \genv, \tenv'}

\nextrulesmall
\irule
{\wfenv\twfholds\bvltyp{\ttC}\nextclause \\ 
	\ttx\notin\genv,\tenv \nextclause \\
	\tenv' = \tenv[\ttx\envgoesto\bvltyp{\ttC}]}
{\genv, \tenv\holds\ttlet\ttstate\ \ttcontrol\ \ttdontgate\ \ttx\ \ttcol\ \bvltyp{\ttC}\ttthen \genv, \tenv'}

\nextrulesmall
\irule
	{\wfenv\twfholds\memtyp{\ttC_1}{\ttC_2}{\ttC_3}\nextclause \\ 
		\ttmemid\notin\genv,\tenv \nextclause \\
		\tenv' = \env[\ttmemid\envgoesto\memtyp{\ttC_1}{\ttC_2}{\ttC_3}]}
	{\genv, \tenv\holds\ttlet\ttstate\ \ttmemid\ \ttcol\ \memtyp{\ttC_1}{\ttC_2}{\ttC_3} \ttthen \genv, \tenv'}

\nextrulesmall
\irule
	{\wfenv\twfholds\memtyp{\ttC_1}{\ttC_2}{\ttC_3}\nextclause \\
		\wfenv\twfholds\labeltyp{\ttC_3}\nextclause \\ \\
		\ttmemid\notin\genv,\tenv \nextclause \\
		\NNEW{\ttx} \notin\genv,\tenv \nextclause \\ \\
		\tenv' = \tenv[\ttmemid\envgoesto\memtyp{\ttC_1}{\ttC_2}{\ttC_3}] \\ 
		\tenv'' = \tenv'[\NNEW{\ttx}\envgoesto\labeltyp{\ttC_3}]}
	{\genv, \tenv\holds\ttlet\ttstate\ \ttmemid\ \ttcol\ \memtyp{\ttC_1}{\ttC_2}{\ttC_3}
		\ \ttwith\ \NNEW{\ttx} \ttthen \genv, \tenv''}

\nextrule
\begin{tabular}{cc}
\irule {~}
	{\genv, \tenv\tholds \eps \ttthen \genv, \tenv}
&
\irule 
{\genv, \tenv\tholds \defop \ttthen \genv', \tenv' \nextclause \\ 
	\genv', \tenv' \tholds \defops \ttthen \genv'', \tenv''}
{\genv,\tenv\tholds \defop \ttsemi\ \defops \ttthen \genv'',\tenv''}
\end{tabular}

\nextrulesmall	
\irule
{\ttxop\notin\genv,\tenv \nextclause \\
	\genv, \env\tholds\hastyp{\tte}{\ttstring}\nextclause \\
	\genv, \env\tholds\ttS \nextclause \\
	\tenv' = \tenv[\ttxop \envgoesto \funtyp{()} {()}] }
{\genv,\tenv\holds\ttdefop\ \ttxop\ \ttinbrace{\tttxt=\tte\ttcomma\ttsem=\ttS}\ttthen\genv, \tenv'}

\nextrulesmall
\irule
{\wfenv\twfholds \left( \funtyp{\sseq{\cbasetypi}}{()} \right)\nextclause \\
	\ttxop\notin\genv,\tenv \nextclause \\ 
	\forall \tti, \cbasetypi \neq \ttstring \wedge \cbasetypi \neq \ttunit \wedge \cbasetypi \neq \cregstyp \nextclause \\
	\tenv' = \tenv[\forall \tti, \ttx_i\envgoesto\cbasetypi]\nextclause \\
	\genv, \env'\tholds\hastyp{\tte}{\ttstring}\nextclause \\
	\genv, \env'\tholds\ttS \nextclause \\
	\tenv'' = \tenv[\ttxop \envgoesto \left( \funtyp{\sseq{\ttx_i\ \ttcol\ \cbasetypi}} {()}\right) ] }
{\genv,\tenv\holds\ttdefop\ \ttxop\ \sseq{\ttx_i\ \ttcol\ \cbasetypi} \ttinbrace{\tttxt=\tte\ttcomma\ttsem=\ttS}\ttthen\genv, \tenv''}
\caption{\casp typing rules for declarations.\label{fig:tr_casp_type_decl}}
\end{figure*}


\mypara{Declarations}
Declarations update the environment.
The form of a typing judgment for a declaration is
$\genv, \tenv \holds \decl \ttthen \genv', \tenv'$, and a judgment for
a list of declarations has the same form, shown in \autoref{fig:tr_casp_type_decl}.
This means that the declaration (or list) is well typed and produces
the new environment on the right.
One can think of the $\holds$ $\ttthen$ as the tail and head of an
arrow with the element transforming the environments labeling the
body.

We impose an additional syntactic restriction on declarations found in
a machine description (as opposed to the additional declarations that
may appear in a specification): they may not use the expression forms that
refer to machine state (registers or memory),
because when defining the machine, there is no specific machine
state to refer to; any references need to be quantified.
(That in turn is not allowed to avoid needing to send these
quantifiers to the SMT solver.)

\begin{figure*}
\textbf{(Machine Typing)}
\nextrulesmall
\irule 
	{\genv_\mathit{builtin}, \tenv_\mathit{builtin}\tholds \decls \ttthen \genv, \tenv \nextclause \\
		\genv, \tenv \tholds \defops \ttthen \genv', \tenv'}
	{\tholds \decls \ttsemi\ \defops \ttthen \genv', \tenv'}
	
\nextrulesmall
\textbf{(Specification Typing)}
\nextrulesmall
\irule 
{\tholds \machine \ttthen \genv, \tenv \nextclause \\
	\genv, \tenv \tholds \decls \ttthen \genv', \tenv' }
{\tholds \machine \ttsemi\ \decls \ttthen \genv', \tenv'}

\nextrulesmall
\begin{tabular}{cc}
	\irule
	{ \forall \tti,\ \genv, \tenv \tholds \hastyp{\ttregidi}{\bvltyp{\ttC_i}}}
	{\typenv \tholds \ttfwrites\ \ttcol\ \sseq{\ttregidi} }
	&
	\irule
	{ \forall \tti,\ \genv, \tenv \tholds \hastyp{\tte_i}{\ttint} \nextclause \\
		\forall \tti,\ \genv, \tenv \tholds \hastyp{\ttmemidi}{\cmemtyp}}
	{\typenv \tholds \ttfmwrites\ \ttcol\ \sseq{\ptrform{\ttmemidi}{\tte_i}} }
\end{tabular}

\nextrulesmall
\begin{tabular}{cc}
	\irule{~}
	{\typenv \tholds \eps}
	&
	\irule
	{\typenv \tholds \ttframe \nextclause \\ \typenv \tholds \ttframes}
	{\typenv \tholds \ttframe\ \ttframes}
\end{tabular}

\nextrulesmall
\irule 
{\tholds \machine \ttsemi\ \decls \ttthen \genv, \tenv \nextclause \\
	\genv, \tenv \tholds \ttframe \nextclause \\
	\genv, \tenv \tholds \hastyp{\ttpre}{\ttbool} \nextclause \\
	\genv, \tenv \tholds \hastyp{\ttpost}{\ttbool} }
{\machine \tholds \decls \ttsemi\ \ttframes \ttsemi\ \ttpre \ttsemi\ \ttpost}
	
\nextrulesmall
\textbf{(Program Typing)}
\nextrulesmall
\begin{tabular}{cc}
	\irule
	{\typenv\tholds\hastyp{\ttxop}{\left( \funtyp{()}{()} \right)}}
	{\typenv\tholds \ttxop}
	&
	\irule
	{\typenv\tholds\hastyp{\ttxop}{\left( \funtyp{\sseq{\cbasetypi}}{()} \right)}\nextclause \\
		\forall \tti,\ \typenv\tholds\hastyp{\tte_i}{\cbasetypi}}
	{\typenv\tholds \ttxop\ {\sseq{\tte_i}}}
\end{tabular}
\nextrulesmall
\begin{tabular}{ccc}
	\irule {~} {\genv, \tenv \tholds \eps}
	&
	\irule 
	{ \genv, \tenv \tholds \itinst \nextclause \\
		\genv, \tenv \tholds \itinsts}
	{\genv, \tenv \tholds \itinst \ttsemi\ \itinsts}
	&
	\irule 
	{\tholds \machine \ttthen \genv, \tenv \nextclause \\
		\genv, \tenv \tholds \program}
	{\machine \tholds \program}
\end{tabular}
\caption{\casp typing rules for machine, specification, and program.\label{fig:tr_casp_type_mach}}
\end{figure*}

\mypara{Machines}
A machine is some declarations followed by some defops, so the typing
rule is just sequencing, shown in \autoref{fig:tr_casp_type_mach}, 
but there is a wrinkle: the initial
environment for the machine is not an input.
$\genv_\mathit{builtin}$ is the (fixed) environment describing the built-in
type aliases.
(Currently there are none.)
$\tenv_\mathit{builtin}$ is the environment describing the types of
built-in variables.
This notionally includes the built-in functions.
(But as mentioned earlier some of them actually have their own typing
rules.)
The form of a typing judgment for a machine is
$\tholds \machine \ttthen \genv, \tenv$.
This means that the machine description is well typed and provides the
environment on the right for use of other constructs that depend on
the machine.
(Specs and programs are only valid relative to a given machine.)

\mypara{Specifications}
For specifications we need two helper rules, shown in
\autoref{fig:tr_casp_type_mach}:
one that applies an
additional list of declarations to a machine, which has the same
form as the judgment on a machine; and one that says that a frame
(modifies list) is well typed, which has the form
$\typenv \tholds \ttframes$.
This lets us write the real rule, which has the form
$\machine \tholds \itspec$
and means that the specification is well typed under the machine description.

\mypara{Programs}
A program is a sequence of calls to instructions.
We need judgments of the form $\typenv \tholds \itinst$ for a single
instruction and also $\typenv \tholds \itinsts$ for the sequence, shown in \autoref{fig:tr_casp_type_mach}.
There are two cases for a single instruction because of a minor glitch
in formulation: because the overbar notation means ``one or more'',
there are two cases in the syntax for instructions, one for zero
operands and one for one or more operands; we need typing rules for
both cases.
Meanwhile the type entered into $\tenv$ for a zero-operand instruction
is unit to unit, not {\eps} to unit, to avoid needing an additional
form for types just for this case.
(Notice that a one-operand instruction may not have type unit to unit
because unit is not allowed as an instruction operand, so the type is
not ambiguous.)
These rules let us write a judgment for a program, which has the form
$\machine \tholds \program$ and means that the program is well typed
relative to the machine.

\mypara{Soundness}
Note that even though we do not check certain things statically, the
type system remains sound: we include the necessary checks and failure
states in the semantics so that evaluation does not get stuck.

\section{\casp{} Semantics}
\label{sec:exec}

This section defines the semantics of \casp.

\begin{figure*}
\textbf{(Expression Semantics)} \\
\nextrulesmall
\begin{tabular}{ccc}
\irule
	{\venv(\ttx)=\ttv}
	{\esreduce{\ttx}{\ttv}}
&
\irule
	{\esreduce{\tte}{\ttr} \nextclause \\
		\venv(\strof{\ttr})=\ttv}
	{\esreduce{\strof{\tte}}{\ttv}}
&
\irule
	{\esreduce{\tte}{\ttr} \nextclause \\
		\strof{\ttr} \notin \venv}
	{\esreduce{\strof{\tte}}{\exprfail}}
\end{tabular}
\nextrulesmall
\irule
	{\forall \tti, \esreduce{\tte_i}{\ttv_i} \nextclause \\ 
		\venv\ttinparen{\ttfuncid} = \{\sseq{\ttx_i}, {\tte}\}\nextclause \\ 
		\esreducevenv{\venv[\forall \tti, \ttx_i \envgoesto \ttv_i]}{\tte}{\opsenv}{\tildev}}
	{\esreduce{\fapp{\ttfuncid}{\sseq{\tte_i}}}{\tildev}}
	
\nextrulesmall
\begin{tabular}{cc}
\irule
	{\esreduce{\tte}{\ttv_1} \nextclause \\
		\tildev_2 = \itunop\ \ttv_1}
	{\esreduce{\itunop\ \tte}{\tildev_2}}
&
\irule
	{\esreduce{\tte_1}{\ttv_1} \ \ \ \ \ 
		\esreduce{\tte_2}{\ttv_2} \ \ \ \ \ 
		\tildev_3 =\ttv_1\ \itbinop\ \ttv_2}
	{\esreduce{\tte_1\ \itbinop\ \tte_2}{\tildev_3}}
\end{tabular}

\nextrulesmall
\begin{tabular}{cc}
\irule
	{\esreduce{\tte}{\bitconst}\nextclause \\   
		\ttC = \ttb_0\ldots\ttb_{\ttC_i}\ldots\ttb_\ttn}
	{\esreduce{\bitidx{\tte}{\ttC_i}}{\ttb_{\ttC_i}}}
&
\irule
	{\esreduce{\tte}{\bitconst}\nextclause \\ 
		\ttC = \ttb_0\ldots\ttb_{\ttC_i}\ldots\ttb_{\ttC_j}\ldots\ttb_\ttn}
	{\esreduce{\bitfield{\tte}{{\ttC_i}}{{\ttC_j}}}{\ttb_{\ttC_i}\ldots\ttb_{\ttC_j}}}
\end{tabular}

\nextrulesmall
\begin{tabular}{cc}
\irule
	{\esreduce{\tte}{\ptrform{\ttmemid}{\ttC}}}
	{\esreduce{\bitidx{\tte}{\_}}{\exprfail}}
&
\irule
	{\esreduce{\tte}{\ptrform{\ttmemid}{\ttC}}}
	{\esreduce{\bitfield{\tte}{{\_}}{{\_}}}{\exprfail}}
\end{tabular}

\nextrulesmall
\irule
	{\esreduce{\tte_1}{\ttv_1} \nextclause \\
		\esreducevenv{\venv[\ttx \envgoesto \ttv_1]}{\tte_2}{\opsenv}{\tildev_2}}
	{\esreduce{\eLet{\ttx}{\cbasetyp}{\tte_1}{\tte_2}}{\tildev_2}}
	
\nextrulesmall
\begin{tabular}{cc}
\irule
	{\esreduce{\tte}{\tttrue}\nextclause \\ 
		\esreduce{\tte_t}{\tildev_t}}
	{\esreduce{\sITE{\tte}{\tte_t}{\_}}{\tildev_t}}
&
\irule
	{\esreduce{\tte}{\ttfalse}\nextclause \\ 
		\esreduce{\tte_f}{\tildev_f}}
	{\esreduce{\sITE{\tte}{\_}{\tte_f}}{\tildev_f}}
\end{tabular}

\nextrulesmall
\begin{tabular}{cc}
\irule
	{\esreduce{\tte}{\ttr}\nextclause \\ 
		\esloclookup{\ttr}}
	{\esreduce{\regderef{\tte}}{\ttv}}
&
\irule
	{\esreduce{\tte}{\ptrform{\ttmemid}{\ttC}}\nextclause \\ 
		\esmemlookup{\ttmemid}{\ttC}{(\ttv, \ttC_l)} }
	{\esreduce{\fetch{\tte}{\ttC_l}}{\ttv}}
\end{tabular}

\nextrulesmall
\irule
	{\esreduce{\tte}{\ptrform{\ttmemid}{\ttC}}\nextclause \\ 
		\esmemlookup{\ttmemid}{\ttC}{(\_, \ttC_m)}\nextclause \\
		\ttC_m \neq \ttC_l}
	{\esreduce{\fetch{\tte}{\ttC_l}}{\exprfail}}

\nextrulesmall
\begin{tabular}{cc}
\irule
	{\esreduce{\tte}{\ptrform{\ttmemid}{\ttC}}\nextclause \\ 
		\ptrform{\ttmemid}{\ttC}\notin\opsmenv}
	{\esreduce{\fetch{\tte}{\ttC_l}}{\exprfail}}
&
\irule
	{\esreduce{\tte}{\bitconst}}
	{\esreduce{\fetch{\tte}{\ttC_l}}{\exprfail}}
\end{tabular}

\nextrulesmall
\irule
	{\esreduce{\tte}{\ttC}}
	{\esreduce{\ptrform{\ttmemid}{\tte}}{\ptrform{\ttmemid}{\ttC}}}

\nextrulesmall
\begin{tabular}{cc}
\NEWW{
\irule
	{\venv(\texttt{EXTBRANCH})=\texttt{ext}}
	{\esreduce{\ebranchto{\ttextid}}{\tttrue}}
}
&
\NEWW{
\irule
	{\venv(\texttt{EXTBRANCH})=\cdot}
	{\esreduce{\ebranchto{\ttextid}}{\ttfalse}}
}
\end{tabular}

\nextrulesmall
\begin{tabular}{cc}
\irule
	{\forall \tti \in(1\ldots\ttk),\ \venv(\ttx_i) = \ttr_i}
	{\esreduce{\ttinbrace{\ttx_1, \ldots, \ttx_k}}{\ttinbrace{\ttr_1, \ldots, \ttr_k}}}
&
\irule
	{\esreduce{\tte}{\ttinbrace{\ttr_1, \ldots, \ttr_{\ttC}}}}
	{\esreduce{\sizeofset{\tte}}{\ttC}}
\end{tabular}

\nextrulesmall
\irule
	{\esreduce{\tte_1}{\ttr} \nextclause \\
		\esreduce{\tte_2}{\ttinbrace{\ttr_1, \ldots, \ttr_k}} \nextclause \\
		\exists \tti \in (1\ldots\ttk), \ttr_{\tti} = \ttr}
	{\esreduce{\setmemberof{\tte_1}{\tte_2}}{\tttrue}}

\nextrulesmall
\irule
	{\esreduce{\tte_1}{\ttr} \nextclause \\
		\esreduce{\tte_2}{\ttinbrace{\ttr_1, \ldots, \ttr_k}} \nextclause \\
		\forall \tti \in (1\ldots\ttk), \ttr_{\tti} \neq \ttr}
	{\esreduce{\setmemberof{\tte_1}{\tte_2}}{\ttfalse}}

\caption{\casp semantics for expressions.\label{fig:tr_casp_sem_expr}}
\end{figure*}

\mypara{Environment}
The execution environment $\venv$ maps \casp variables \ttx to values
\ttv.
\NNEW{For labels on memory regions, each label
maps to a pointer that points to the base (offset 0) of the region
associated with the label.}
However, we take advantage of the polymorphism and dynamic
typing of paper rules to also store the following in the same
environment:
\begin{itemize}
\item \ttfuncid (function names) map to pairs $\{\sseq{\ttx_i},
{\tte}\}$, which give the list of argument names and the body for
functions.
\item \ttprocid (procedure names) map to pairs $\{\sseq{\ttx_i},
{\ttS}\}$, which give the list of argument names and the body for
procedures.
\item \ttxop (operation/instruction names) map to triples $\{\sseq{\ttx_i},
\tte, {\ttS}\}$, which give the list of argument names, the expression
for the text form, and the body for operations.
\item \strof{\ttr} (the form for the text version of a register) maps
to a value.
\item \NEWW{The word \texttt{EXTBRANCH} maps to a branch state, which
must be either \texttt{ext} or $\cdot$.}
This reports whether, after executing a program, it branched to the
external label or not.
\end{itemize}
Since identifiers are not allowed to overlap in well-typed programs,
and register identities are not strings at all, \NEWW{and \texttt{EXTBRANCH}
is
reserved,} this usage creates no
conflicts.

Note that \ttmemid, \tttypeid, and \ttxmoduleid do not appear in
$\venv$ as these require no translation/lookup at runtime.

\mypara{Machine State}
In addition to the execution environment, we also need a representation
of machine state.
We define two stores, one for registers and one for memory.
The register store $\opsrenv$ maps registers \ttr to values \ttv.
The memory store $\opsmenv$ is more complicated: it maps pairs
$\ptrform{\ttmemid}{\ttC}$ (that is, pointer literals) to pairs
$(\ttv, \ttC_l)$, where \ttv is the value stored at that location and
$\ttC_l$ is the bit width.
The bit widths of memory regions are invariant, both across the region
when they are declared and also over time.
They are used to check the access widths appearing in fetch and store
operations.
Also note that new entries cannot be created in either the register
store or the memory store, as real hardware does not permit such
actions.
The values stored in registers and memory regions are restricted by
the typing rules to bitvectors (whether constants or pointers) of the
appropriate width.

Notice that stepping through the declarations does \emph{not}
initialize the machine state.
We want to reason about executions over ranges of possible starting
machine states; so instead we provide a judgment that uses the typing
environments to restricts the stores to forms consistent with the
declarations.
This is discussed further below.

\mypara{Expressions}
We describe expressions with a large-step operational semantics, shown in \autoref{fig:tr_casp_sem_expr}.
The form of an expression semantic judgment is:
$\esreduce{\tte}{\ttv}$,
which means that given the environment $\venv$ and the machine
state $\opsenv$, the expression \tte evaluates to the value \ttv.
Expressions may refer to the machine state, but not modify it.
Expressions can fail; in addition to the explicit failure cases seen,
some of the operators and built-in functions can fail.
For example, as mentioned earlier, attempting bitvector arithmetic
other than addition and subtraction on pointers will fail.
Furthermore, division by zero fails.

Note that we currently do not statically check (in the typing rules)
that the \strof{} form is present for every register or that it is
defined for registers on which it is used.
Thus we have an explicit failure rule for when no matching declaration
has been seen.
We also have failure rules for bad fetch operations: if the length
annotation is wrong, if the pointer is not in the machine state
(this covers both unaligned accesses and out of bounds accesses),
or if the value used is not a pointer.
Similarly, we have failure rules for when bit indexing/slicing a pointer. 
We do not, conversely, need explicit failure checks or rules for the
bit indexes in the bit extraction/slicing constructs as they are
statically checked.

Also note that we include in the semantics the obvious failure
propagation rules for when subexpressions fail.
We do not show these explicitly as they are not particularly
interesting or informative; however, note that the $\&\&$ and $||$
logical operators short-cut left to right.


\begin{figure*}
\textbf{(Statement Semantics)}\\
\nextrulesmall
\irule
	{\osreduce{\ttS_1}{\opsenv}{\ttskip}{\opsenvB}\nextclause \\
		\osreduce{\ttS_2}{\opsenvB}{\ttS_2'}{\opsenvC}}
	{\osreduce{\ttS_1\ttsemi\ \ttS_2}{\opsenv}{\ttS_2'}{\opsenvC}}

\nextrulesmall
\irule
{\forall \tti, \esreduce{\tte_i}{\ttv_i} \nextclause \\  	
	\venv\ttinparen{\ttprocid} = \{\sseq{\ttx_i},\ttS\}\nextclause \\ 
	\osreducetenv{\venv[\forall \tti, \ttx_i \envgoesto \ttv_i]}{\ttS}{\opsenv}{\ttS'}{\opsenvB}}
{\osreduce{\papp{\ttprocid}{\sseq{\tte_i}}}{\opsenv}{\ttS'}{\opsenvB}}

\nextrulesmall
\irule
	{\esreduce{\tte}{\ttv} \nextclause \\  
		\osreducetenv{\venv[\ttx \envgoesto \ttv]}{\ttS}{\opsenv}{\ttS'}{\opsenvB}}
	{\osreduce{\sLet{\ttx}{\cbasetyp}{\tte}{\ttS}}{\opsenv}{\ttS'}{\opsenvB}}

\nextrulesmall
\irule
	{\forall \tti \in (\ttC_1, \ttC_1+1,\ \ldots, \ttC_2),\ 
		\osreducetenv{\venv[\ttx \envgoesto \tti]}{\ttS}{\opsenvafter{i}{i}}{\ttskip}{\opsenvafter{i+1}{i+1}}}
	{\osreduce{\sFor{\ttx}{\ttC_1}{\ttC_2}{\ttS}}{\opsenvafter{\ttC_1}{\\tC_1}}{\ttskip}{\opsenvafter{\ttC_2+1}{\ttC_2+1}}}

\nextrulesmall
\irule
	{\esreduce{\tte}{\tttrue}\nextclause \\
		\osreduce{\ttS}{\opsenv}{\ttS_t}{\opsenvafter{t}{t}}}
	{\osreduce{\sITE{\tte}{\ttS}{\_}}{\opsenv}{\ttS_t}{\opsenvafter{t}{t}}}
	
\nextrulesmall
\irule
	{\esreduce{\tte}{\ttfalse}\nextclause \\
		\osreduce{\ttS}{\opsenv}{\ttS_f}{\opsenvafter{f}{f}}}
	{\osreduce{\sITE{\tte}{\_}{\ttS}}{\opsenv}{\ttS_f}{\opsenvafter{f}{f}}}

\nextrulesmall
\irule
	{\esreduce{\tte_1}{\ttr} \nextclause \\ 
		\ttr \in \opsrenv \nextclause \\
		\esreduce{\tte_2}{\ttv} \nextclause \\ 
		\opsrenv' = \opsrenv[\ttr\envgoesto\ttv]}
	{\osreduce{\slAssign{\tte_1}{\tte_2}}{\opsenv}{\ttskip}{\opsrenv',\opsmenv}}

\nextrulesmall
\irule
	{\esreduce{\tte_1}{\ptrform{\ttmemid}{\ttC}}\nextclause \\ 
		\esmemlookup{\ttmemid}{\ttC}{(\_, \ttC_l)} \nextclause \\
		\esreduce{\tte_2}{\ttv} \nextclause \\
		\opsmenv' = \opsmenv[ {\ptrform{\ttmemid}{\ttC}} \envgoesto(\ttv, \ttC_l)]}
	{\osreduce{\smAssign{\tte_1, \ttC_l}{\tte_2}}{\opsenv}{\ttskip}{\opsrenv,\opsmenv'}}

\nextrulesmall
\irule
	{\esreduce{\tte_1}{\ptrform{\ttmemid}{\ttC}}\nextclause \\ 
		\esmemlookup{\ttmemid}{\ttC}{(\_, \ttC_{m})} \nextclause \\
		\ttC_{m} \neq \ttC_{l}}
	{\osreduce{\smAssign{\tte_1, \ttC_l}{\tte_2}}{\opsenv}{\bferr}{\opsrenv,\opsmenv}}

\nextrulesmall
\begin{tabular}{cc}
\irule
	{\esreduce{\tte_1}{\ptrform{\ttmemid}{\ttC}}\nextclause \\ 
		{\ptrform{\ttmemid}{\ttC}} \notin \opsmenv}
	{\osreduce{\smAssign{\tte_1, \ttC_l}{\tte_2}}{\opsenv}{\bferr}{\opsrenv,\opsmenv}}
&
\irule
	{\esreduce{\tte_1}{\bitconst}}
	{\osreduce{\smAssign{\tte_1, \ttC_l}{\tte_2}}{\opsenv}{\bferr}{\opsrenv,\opsmenv}}
\end{tabular}

\nextrulesmall
\begin{tabular}{cc}
\irule
	{\esreduce{\tte}{\tttrue}}
	{\osreduce{\ttassert(\tte)}{\opsenv}{\ttskip}{\opsenv}}
&
\irule
	{\esreduce{\tte}{\ttfalse}}
	{\osreduce{\ttassert(\tte)}{\opsenv}{\bferr}{\opsenv}}
\end{tabular}

\nextrulesmall
\begin{tabular}{cc}
\NEW{
\irule
	{\esreduce{\tte}{0}}
	{\osreduce{\sBRANCH{\tte}}{\opsenv}{\ttskip}{\opsenv}}
}
&
\NEW{
\irule
	{\esreduce{\tte}{\bitconst} \nextclause \\
	 \texttt{0x00} < \bitconst < \texttt{0xff}}
	{\osreduceBR{\sBRANCH{\tte}}{\opsenv}{\ttskip}{\opsenv}{\bitconst}}
}
\end{tabular}

\nextrulesmall
\NEW{
\irule
	{\esreduce{\tte}{\texttt{0xff}}}
	{\osreduceBR{\sBRANCH{\tte}}{\opsenv}{\ttskip}{\opsenv}{\texttt{ext}}}
}
\caption{\casp semantics for statements.\label{fig:tr_casp_sem_stmt}}
\end{figure*}

\mypara{Statements}
Unlike expressions, statements can change machine state.
Thus, the form of a machine state semantics judgment (also large step)
is
$\osreduceBR{\ttS}{\opsenv}{\ttS'}{\opsenvB}{\xi}$, shown in \autoref{fig:tr_casp_sem_stmt}.
This means that the statement \ttS evaluates to the irreducible
statement \ttS' (which must be either \ttskip or \bferr) and in the
course of doing so changes the machine state from \opsenv to \opsenvB,
\NEW{and produces a branching state $\xi$, which can be either an 8-bit
bitvector, the reserved value \texttt{ext}, or a dot ($\cdot$).}
As with expressions, statements can fail.
Explicit failure rules are shown for bad stores (corresponding to the
cases for bad fetches) and for a failed assertions.
We also similarly include, but do not show, the obvious failure
propagation rules for cases where sub-statements, or expressions
within statements, fail.

\begin{figure*}
\textbf{(Declaration Semantics)}\\
\nextrulesmall
\begin{tabular}{cc}
\irule
   {~}
   {\SEMdeclAA{\eps}}
&
\irule
   {\SEMdeclAB{\ttdecl} \nextclause \\
      \SEMdeclBC{\ttdecls}}
   {\SEMdeclAC{\ttdecl \ttsemi\ \ttdecls}}
\end{tabular}

\nextrulesmall
\begin{tabular}{cc}
\irule
   {~}
   {\SEMdeclAA{\tttype\ \tttypeid = \cbasetyp}}
&
\irule
   {\esreduce{\tte}{\ttv}}
   {\SEMdeclAx{\ttlet\ \ttx\ \ttcol\ \cbasetyp = \tte}{\ttx \envgoesto \ttv}}
\end{tabular}

\nextrulesmall
\irule
   {\esreduce{\ttx}{\ttr} \nextclause \\ 
   		\esreduce{\tte}{\ttv}}
   {\SEMdeclAx{\ttlet\ \strof{\ttx} = e}{\strof{\ttr} \envgoesto \ttv}}

\nextrulesmall
\begin{tabular}{cc}
\irule
   {\venv' = \venv[\ttfuncid \envgoesto \{\sseq{\ttx_i}, \tte\}]}
   {\SEMdeclAB{\ttdef\ \ttfuncid\ \funtyp{\sseq{\ttx_i\ \ttcol\ \cbasetypi}}{\cbasetyp} = \tte}}
&
\irule
{\venv' = \venv[\ttprocid \envgoesto \{\sseq{\ttx_i}, \ttS\}]}
{\SEMdeclAB{\ttproc\ \ttprocid\ \proctyp{\sseq{\ttx_i\ \ttcol\ \cbasetypi}} = \ttS}}
\end{tabular}

\nextrulesmall
\begin{tabular}{cc}
\irule
   {\venv' = \venv[\ttx \envgoesto \ttr] \nextclause \\	
   		\ttr\ \mathrm{fresh}}
   {\SEMdeclAB{\ttlet\ttstate\ \ttx\ \ttcol\ \cregtyp}}
&
\irule
{\venv' = \venv[\ttx \envgoesto \ttr] \nextclause \\	
	\ttr\ \mathrm{fresh}}
{\SEMdeclAB{\ttlet\ttstate\ \ttcontrol\ \ttx\ \ttcol\ \cregtyp}}
\end{tabular}

\nextrulesmall
\irule
{\venv' = \venv[\ttx \envgoesto \ttr] \nextclause \\	
	\ttr\ \mathrm{fresh}}
{\SEMdeclAB{\ttlet\ttstate\ \ttcontrol\ \ttdontgate\ \ttx\ \ttcol\ \cregtyp}}

\nextrulesmall
\begin{tabular}{cc}
\irule
   {~}
   {\SEMdeclAA{\ttlet\ttstate\ \ttmemid\ \ttcol\ \cmemtyp}}
&
\irule
   {\venv' = \venv[\NNEW{\ttx} \envgoesto \ptrform{\ttmemid}{0}] \nextclause \\	}
   {\SEMdeclAB{\ttlet\ttstate\ \ttmemid\ \ttcol\ \cmemtyp\ \ttwith\ \NNEW{\ttx}}}
\end{tabular}

\nextrule
\begin{tabular}{cc}
\irule
   {~}
   {\SEMdeclAA{\eps}}
&
\irule
   {\SEMdeclAB{\defop} \nextclause \\
      \SEMdeclBC{\defops}}
   {\SEMdeclAB{\defop \ttsemi\ \defops}}
\end{tabular}
   
\nextrulesmall
\irule
{ \venv' = \venv[\ttxop \envgoesto \{[~], \tte, S\}]}
{\SEMdeclAB{\ttdefop\ \ttxop\ \ttinbrace{\tttxt=\tte\ttcomma\ttsem=\ttS}}}

\nextrulesmall
\irule
   { \venv' = \venv[\ttxop \envgoesto \{\sseq{x_i}, \tte, S\}]}
   {\SEMdeclAB{\ttdefop\ \ttxop\ \sseq{\ttx_i\ \ttcol\ \cbasetypi}\ \ttinbrace{\tttxt=\tte\ttcomma\ttsem=\ttS}}}
\caption{\casp semantics for declarations.\label{fig:tr_casp_sem_decl}}
\end{figure*}

\mypara{Declarations}
The semantics for declarations have judgments of the form
$\SEMdeclAB{\ttdecl}$, shown in \autoref{fig:tr_casp_sem_decl}.
This means that the given declaration updates $\venv$ as shown.
As stated above, we do not \emph{initialize} the machine state while
handling declarations; this instead allows us to work with arbitrary
(or universally quantified) machine states afterwards.
However, because the let-binding declaration evaluates an expression,
it potentially needs \emph{access} to a machine state.
Consequently we write the rules so they accept a machine state as input,
but do not update it.
In the case of machine descriptions, where there is no machine state,
we pass empty environments; let declarations in machine descriptions
are not allowed to reference machine state.
In the case of the additional declarations that accompany a
specification, we pass in the initial machine state; this allows
values from the initial machine state to be globally bound so they can
be referred to in the postcondition.

We give first the rules for a list of declarations, then the rules for the various declarations, then the
rules for a list of operation definitions and a rule for a single operation definition.
Note that several of the declarations do not update $\venv$, and
nothing is placed in $\venv$ for memory regions.
For registers, only the mapping of the identifier to its
underlying register \ttr is entered; nothing for \ttr is inserted.

\begin{figure}
\textbf{(Machine Semantics)}\\
\nextrule
\irule
   {\SEMZZA{\decls} \nextclause \\
      \SEMmachAB{\defops}}
   {\SEMmachB{\decls \ttsemi\ \defops}}
\caption{\casp semantics for machines.\label{fig:tr_casp_sem_mach}}
\end{figure}

\mypara{Machines}
As with the typing rules, the semantics rule for a whole machine
description integrates the initial environment and gives a judgment of
the form $\SEMmachB{\machine}$, shown in \autoref{fig:tr_casp_sem_mach}.
We also include a comparable form that includes additional
declarations, as it will be used below.

\mypara{Programs}
Instructions update the machine state, and we chose to represent
programs as lists of instructions rather than having dummy
instruction forms for skip and sequence.
Consequently the form of the judgments is slightly different, and
there are several of them, shown in \autoref{fig:tr_casp_sem_prog}.

First, we can execute an individual instruction using the form
$\psreduce{\itinst}{\opsenv}{\opsenvB, \NEW{\xi}}$,
meaning that the instruction executes and updates the machine state
\opsenv to \opsenvB, \NEW{producing the branching state $\xi$}.
Then, a list of instructions executes using the form
\psreduce{\itinsts, \NEW{\xi}}{\opsenv}{\opsenvB, \itinsts', \NEW{\xi'}},
which means that the list steps to a new list and updates \NEW{both} the
machine state \NEW{and the branching state}.
When the instruction list runs out, these reduce to a shorter form
via a judgment of the form
\psreduce{\itinsts, \NEW{\xi}}{\opsenv}{\opsenvB},
which discards the instructions and branching state and produces an
output machine state.
That in turn allows us to draw conclusions of the form
$\ttmach \oholds(\opsenv,\program)\goesto(\opsrenv', \opsmenv')$,
which means that a machine with the initial state $\opsenv$ executes
the program to produce the new machine state $(\opsrenv', \opsmenv')$.


Note that there are two versions of the judgment for instructions,
one specialized for no arguments/operands.
Instructions with no operands are declared as taking unit, but
invoked with empty operands (not with unit) to correspond to the way
assembly languages normally work.

We include a final judgment of the form 
$\ttmach \oholds(\opsenv,\program)\goesto(\opsrenv', \opsmenv')$
that puts the machine on the left-hand side of the turnstile.
It means that under a given machine the program maps \opsenv to
\opsenvB.
There is a limitation in the way we have formulated programs and the
rules for programs, which is that there is no easy way to represent
failure.
(Failure in this might represent triggering an exception and stopping
execution, which we do not model, or invoking ``unpredictable'' or
``undefined'' behavior in the processor and transitioning to an
arbitrary unknown machine state.)

The intended behavior is that a program that fails during execution
(that is, the body of one of its instructions steps to \bferr) enters
a state where no postcondition can evaluate to \tttrue.
We have decided for the moment that working this explicitly into the
formalism would result in a lot of complication and obscuration
without providing any useful information.

\begin{figure*}
\textbf{(Program Semantics)}\\
\nextrulesmall
\irule
   {\venv(\ttxop) = \{[~], \_, S)\} \nextclause \\
   	\osreduceBR{\ttS}{\opsenv}{\ttskip}{\opsenvB}{\xi}}
   {\psreduce{\ttxop}{\opsenv}{\opsenvB, \NEW{\xi}}}

\nextrulesmall
\irule
   {\forall \tti, \esreduce{\tte_i}{\ttv_i} \nextclause \\
      \venv(\ttxop) = \{\sseq{\ttx_i}, \_, S\}\nextclause \\
      \venv' = \venv[\forall i, \ttx_i \envgoesto \ttv_i] \nextclause \\
      \osreduceafterBR{\ttS}{\opsenv}{\ttskip}{\opsenvB}{\xi}}
  {\psreduce{\ttxop\ \sseq{\tte_i}}{\opsenv}{\opsenvB, \NEW{\xi}}}

\nextrulesmall
\begin{tabular}{cc}
\irule
	{\psreduce{\itinst}{\opsenv}{\opsenvB, \NEW{\xi}}}
	{\psreduce{\itinst \ttsemi\ \itinsts, \NEW{\cdot}}{\opsenv}{\opsenvB,
	\NEW{\itinsts, \xi}}}
&
\NEW{
	\irule{~}
	{\psreduce{\itinst \ttsemi\ \itinsts, \texttt{0x01}}{\opsenv}{\opsenv,
			\itinsts, \cdot}}
}
\end{tabular}
\nextrulesmall
\begin{tabular}{cc}
\NEW{
\irule
	{\bitconst > \texttt{0x01}}
	{\psreduce{\itinst \ttsemi\ \itinsts, \bitconst}{\opsenv}{\opsenv,
	\itinsts, \bitconst\ \bitsub\ \texttt{0x01}}}
}
&
\NEW{
\irule
  {~}
  {\psreduce{\_, \texttt{ext}}{\opsenv}{\opsenv,\NEWW{\eps,\texttt{ext}}}}
}
\end{tabular}

\nextrulesmall
\irule
   {\SEMmachA{\ttmach} \nextclause \\	
      \psreducemulti{\program, \NEW{\cdot}}{\opsenv}{\opsrenv',\opsmenv',\NEWW{\eps,\xi}}}
   {\ttmach \oholds(\opsenv,\program)\goesto(\opsrenv', \opsmenv',\NEWW{\xi})}

\caption{\casp semantics for programs.\label{fig:tr_casp_sem_prog}}
\end{figure*}

\begin{figure*}
\textbf{(Specification Semantics)}\\
\nextrulesmall 
\irule
	{\forall \tti, \esreduce{\ttregidi}{\ttr_i} \nextclause \\
		\forall \ttr \notin \{\sseq{\ttr_i}\}, \opsrenv (\ttr) = \opsrenv' (\ttr)}
	{\venv, \opsrenv, \opsmenv, \opsrenv', \opsmenv' \holds \ttfwrites\ \ttcol\ \sseq{\ttregidi} }

\nextrulesmall 
\irule
{\forall \tti, \esreduce{\ptrform{\ttmemidi}{\tte_i}}{\ptrform{\ttmemidi}{\ttC_i}} \nextclause \\
	\forall \ttmemid, \ttC, \ptrform{\ttmemid}{\ttC} \notin \{\sseq{\ptrform{\ttmemidi}{\ttC_i}}\}, \\
	\opsmenv (\ptrform{\ttmemid}{\ttC}) = \opsmenv' (\ptrform{\ttmemid}{\ttC})}
{\venv, \opsrenv, \opsmenv, \opsrenv', \opsmenv' \holds \ttfmwrites\ \ttcol\ \sseq{\ptrform{\ttmemidi}{\tte_i}} }

\nextrulesmall
\begin{tabular}{cc}
	\irule{~}
	{\venv, \opsrenv, \opsmenv, \opsrenv', \opsmenv' \holds \eps}
	&
	\irule
	{\venv, \opsrenv, \opsmenv, \opsrenv', \opsmenv' \holds \ttframe
		\nextclause \\
		\venv, \opsrenv, \opsmenv, \opsrenv', \opsmenv' \holds
		\ttframes
	}
	{
		\venv, \opsrenv, \opsmenv, \opsrenv', \opsmenv' \holds
		\ttframe\ \ttframes
	}
\end{tabular}

\nextrulesmall
\irule
{	\tholds \machine \ttsemi\ \ttdecls \ttthen \genv, \tenv \nextclause \\ 
  \SEMmachA{\machine} \nextclause \\ 
  \forall \opsenv,\ \left( \genv, \tenv, \venv \tholds \opsenv \right) \implies \\
  \SEMAAB{\decls} \implies \\
  \esreduceB{\ttpre}{\tttrue} \implies \\
  \forall \opsrenv', \opsmenv',
  \left( \psreducemultiB{\program,\NEWW{\cdot}}{\opsrenv,\opsmenv}
      {\opsrenv',\opsmenv',\NEWW{\eps,\xi}} \right) \implies \\
  ( \ensm{\venv'\NEWW{[\texttt{EXTBRANCH} \envgoesto \xi]}
      \oholds(\opsrenv', \opsmenv',\ttpost) \egoesto \tttrue} \ \wedge \ 
  \venv', \opsrenv, \opsmenv, \opsrenv', \opsmenv' \holds \ttframes )
}
{ \machine, (\ttdecls\ttsemi\ \ttframe\ttsemi\ \ttpre\ttsemi\ \ttpost) \holds \program}

\caption{\casp semantics for specifications.\label{fig:tr_casp_sem_spec}}
\end{figure*}

\mypara{Specifications}
For specifications we need three judgments, shown in \autoref{fig:tr_casp_sem_spec}: 
the first two state what the
\ttfwrites and \ttfmwrites clauses mean, respectively 
(they are properties on initial and final
register and memory states),  
and the last one says what it means for a program to
satisfy a specification.
Note that the \ttfwrites and \ttfmwrites rules as shown are slightly misleading, because
the register and pointer list provided in the input specification is implicitly
augmented with all registers and pointers mentioned in the postcondition before it
gets to this point.

\begin{figure*}  
\textbf{(Machine State Validity)}
\nextrulesmall
\irule 
{
	\left( \forall \ttx,\ \genv, \tenv \tholds \ttx\ \ttcol\ \bvltyp{\ttC} \wedge \venv (\ttx) = \ttr \right) \Leftrightarrow 
	\left( \exists \ttv,\ \opsrenv (\ttr) = \ttv \wedge \genv, \tenv \tholds \ttv\ \ttcol\ \bvaltyp{\ttC} \right) }
{\typenv, \venv \tholds \opsrenv}

\nextrulesmall
\irule 
{
	\forall \ttmemid,\ \genv, \tenv \tholds \ttmemid\ \ttcol\ \memtyp{\ttC_1}{\ttC_2}{\ttC_3} \Leftrightarrow \\
	( \forall\tti \in \{ 0, \ttC_1 / 8, \dots, (\ttC_2 - 1) * \ttC_1 / 8 \}, \ 
	\exists \ttv,\ \opsmenv (\ttmemid, \tti) = (\ttv, \ttC_1) \wedge \genv, \tenv \tholds \ttv\ \ttcol\ \bvaltyp{\ttC_1} )\ \wedge\\
	( \forall\tti \notin \{ 0, \ttC_1 / 8, \dots, (\ttC_2 - 1) * \ttC_1 / 8 \}, \ 
	\esreduce{\ptrform{\ttmemid}{\tti}}{\exprfail}) }
{\typenv \tholds \opsmenv}

\nextrulesmall
\irule 
{
	\typenv, \venv  \tholds \opsrenv \nextclause \\
	\typenv \tholds \opsmenv \nextclause
}
{\typenv, \venv  \tholds \opsenv}
\caption{\casp machine state validity.\label{fig:tr_casp_mach_valid}}
\end{figure*}

\mypara{Machine State Validity}
As discussed above, we do not initialize the machine state while
processing declarations.
Instead we treat the starting machine state as an input (e.g., in the
final judgment about programs) or quantify it universally as in the
specification judgment.
This requires that we have a predicate to reject
machine states that do not match the machine description.
The validity judgment has the form $\typenv, \venv \tholds
\opsrenv$, shown in \autoref{fig:tr_casp_mach_valid}, 
and correspondingly for $\opsmenv$ (except without $\venv$)
and then for $\opsenv$ (both stores at once).
This means that the given stores match the given environments.

We use this with the typing environments that come from
both the machine description and the
additional declarations arising from a specification.
In the case of registers we need access to $\venv$ to
handle the names of registers.
We do \emph{not} use the $\venv$ generated from the additional
declarations in a specification; this avoids circularity.
This is acceptable, because specifications are not allowed to
define new registers.
For memory regions we need to enumerate the valid offsets for the
region (note the literal 8 that hardwires 8-bit bytes) and check the
cell width.

\mypara{Branching}
\NEW{
To handle branching, we allow statements to produce a branching state,
which indicates the number of instructions to skip over.
Normally this is $\cdot$, which means none and has no effect; however,
when an instruction body produces something else we use it to branch.
A nonzero bitvector results in skipping the indicated number of
instructions; the out-of-band additional value \texttt{ext} causes a
branch to the external label.
The magic number used to select the external label appears only in one
of the statement rules; beyond that point we use \texttt{ext}
explicitly.
}

\NEWW{
The program rule in Figure \ref{fig:tr_casp_sem_spec} inserts the
branch state produced by the program into $\venv$, where it provides a
value for the \texttt{branchto} predicate found in the postcondition.
This allows specifications to enforce the external branching behavior.
}

\newpage

\section{\ale Overview}
\label{sec:ale_lang}
This section describes \ale, our specification language for
writing abstracted machine-independent specifications of low-level
code.

\ale specifications are abstractions of machine-level \casp
specifications; we say that \casp constructs are \emph{lifted} into
\ale and \ale constructs are \emph{lowered} into \casp.
\ale is only for specifications, so there are no statements, no
updates, and no notion of instructions or programs.
We refer to the single synthesis problem in one \ale specification as a \emph{block}.

\mypara{Notation}
We use the following metavariables:

\begin{tabular}{llr}
\ttx, \tty, \ttz & Program variables (binders) \\
\ttr            & Registers (abstract) \\
\ttC            & Integer constants (written in decimal) \\
\bitconst       & Bitvector constants (written in binary) \\
\ttN            & Symbolic integer constants \\
\atyp           & Types \\
\ttv            & Values \\
\tte            & Expressions \\
\tti, \ttj      & Rule-level integers \\
\end{tabular}

(Other constructions are referred to with longer names.)

As noted previously, \ale types and expressions should be
considered distinct from \casp ones (even where they correspond
directly).
We use the same letters in the hopes that this will cause less
confusion (even in the definition of the translation) than picking an
entirely different set of letters for \ale.

\mypara{Identifiers and Variables}
In \ale there are \NNEW{five} syntactic categories of identifiers:
As in \casp, \ttmemid name memory regions.
\ttfuncid name functions, and \tttypeid are type aliases.
\ttxmoduleid name \casp \lowering modules, 
which are used to instantiate the abstract elements and conduct \ale{}--\casp translation for synthesis.
Other \ttx are ordinary variables that range over other things, and
may be presumed to not range over the above reserved categories.
All variables are immutable, in the sense that their values do not change once
bound.

\begin{figure*}
\centering
\begin{minipage}{0.48\textwidth}
\begin{align*}
	& & \textbf{(\ale Symbolic Constants)} \\
	& \ttN \bnfeq & \ttC \bnfor \ttx \\
	& & \\
	& & \textbf{(\ale Types)} \\
	& \atyp \bnfeq & \aprimitivetyp \bnfor \amemorytyp \bnfor \afunctiontyp \\
	& & \\
	& \aprimitivetyp \bnfeq & \ttint \bnfor \ttbool \bnfor \tttypeid \\
	& \bnfor & \ttN\ \ttwvec \bnfor \ttN\ \ttwptr \bnfor \ttN\ \abvltyp \\
	& \bnfor & \ttN\ \ttlabel \bnfor \aregsettyp\\
	& \aregsettyp \bnfeq & \ttN\ \abvstyp\\
	& & \\
	& \amemorytyp \bnfeq & \memtyp{\ttN_1}{\ttN_2}{\ttN_3} \\
	& \afunctiontyp \bnfeq & \afuntyp \\
	& & \\
& & \textbf{(\ale Values)} \\
& \ttv \bnfeq & \tttrue \bnfor \ttfalse \bnfor \ttC \bnfor \bitconst \bnfor \ptrform{\ttmemid}{\ttC} \\
& \bnfor & \ttassertfalse
\end{align*}
\end{minipage}
\hfill\vline\hfill
\begin{minipage}{0.49\textwidth}
\begin{align*}
	& & \textbf{(\ale Operators)} \\
	& \itunop \bnfeq & -  \bnfor \bitsub\ \bnfor \neg\ \bnfor \bitnot\\ 
	& \itbinop \bnfeq &  =\ \bnfor\ \neq\ \bnfor\ + \bnfor - \bnfor * \bnfor / \ \bnfor\ <\ \bnfor\ <=\ \bnfor\ >\ \bnfor\ >= \\
	& \bnfor & \&\& \bnfor || \bnfor \xor \\
	& \bnfor & >> \bnfor {>>}_S \bnfor << \bnfor \bitand \bnfor \bitor \bnfor \bitxor \\
	& \bnfor & \bitadd \bnfor \bitsub \bnfor \bitmul \bnfor \bitdiv  \\
	& \bnfor & \bitlt \bnfor \bitle \bnfor \bitgt \bnfor \bitge  \\
	& \bnfor & \bitslt \bnfor \bitsle \bnfor \bitsgt \bnfor \bitsge\\
	& \bnfor & \cup \bnfor \cap \bnfor \subseteq \bnfor \setminus \\
	& & \\
	& & \textbf{(\ale Expressions)} \\
	& \ttae \bnfeq & \ttv \bnfor \ttx \\
	& \bnfor & \fapp{\ttfuncid}{\sseq{\ttae}} \\ 
	& \bnfor & \itunop\ \ttae \\
	& \bnfor & \ttae_1\ \itbinop\ \ttae_2 \\
	& \bnfor & \bitidx{\ttae}{\ttC} \bnfor \bitfield{\ttae}{\ttC_1}{\ttC_2} \\ 
	& \bnfor & \eLet{\ttx}{\cbasetyp}{\ttae_1}{\ttae_2} \\ 
	& \bnfor & \sITE{\ttae_1}{\ttae_2}{\ttae_3} \\ 
	& \bnfor & \ptrform{\ttmemid}{\ttae} \\ 
	& \bnfor &  \regderef{\ttae}  \bnfor \fetch{\ttae}{\ttN} \\
	& \bnfor & \NEW{\texttt{\ebranchto{sym}}} \\
	& \bnfor & \setlit \\
	& \bnfor & \sizeofset{\ttae} \bnfor \setmemberof{\ttae_1}{\ttae_2} 
\end{align*}
\end{minipage}
\caption{\ale symbolic constants, types, values, operators and expressions.\label{fig:tr_ale_types1}}
\end{figure*}

\mypara{Symbolic Constants}
In \ale symbolic constants \ttN are permitted to
occur in some places where only integer constants are allowed in the
corresponding \casp constructions.
In particular, the bit sizes associated with types (and the lengths of
memory regions, which are functionally similar) may be given as
symbolic values \ttx instead of integer constants.
These must be bound to integer constants either directly in the
\ale spec, in the \casp \lowering, or by the \casp machine
description.
This allows the concrete sizes of bitvectors to vary depending on the
machine architecture.


\mypara{Types}
As in \casp, \ale types are divided syntactically into base types
and others, shown in \autoref{fig:tr_ale_types1}.
The chief difference from \casp is that bit widths (and the lengths of
memory regions) can be symbolic constants.
However, an additional difference is that pointers (\ttwptr) are
distinguished from plain bitvectors (\ttwvec).
This is reasonably possible in \ale, because it need not reason
about the progression of values through machine registers, only before
and after states.
Strings and unit are also absent, as they are not needed for
specifications.



\mypara{Values and Expressions}
The values in \ale correspond directly to the values in \casp as do
operators and most expressions, shown in \autoref{fig:tr_ale_types1}.
Note that the width argument of fetch can be a symbolic size.


\begin{figure*}
\centering
\begin{minipage}{0.54\textwidth}
\begin{align*}
	& & \textbf{(\ale Declarations)} \\
	& \ttdecls \bnfeq & \eps \bnfor \ttdecl\ttsemi\ \ttdecls \\
	& \ttdecl \bnfeq & \ttareq\ \ttatype\ \tttypeid \\ 
	& \bnfor & \ttareq\ \ttaval\ \hastyp{\ttx}{\aprimitivetyp} \\ 
	& \bnfor & \ttareq\ \ttafunc\ \hastyp{\ttfuncid}{\afunctiontyp} \\ 
	& \bnfor & \ttaprov\tsp\ttatype\tsp\tttypeid = \atyp\\
	& \bnfor & \ttaprov\tsp\ttaval\tsp\hastyp{\ttx}{\aprimitivetyp} = \ttae \\
	& \bnfor & \ttaprov\ \ttafunc\ \hastyp{\ttfuncid}{\funtyp{\sseq{\ttx_i\ \ttcol\ \cbasetypi}}{\cbasetyp}} = \ttae \\
	& \bnfor & \ttaregion\ \hastyp{\ttmemid}{\amemorytyp}   \\
	& \bnfor & \ttaregion\ \hastyp{\ttmemid}{\amemorytyp}\ \ttwith\ \NNEW{\ttx} \\
	& \bnfor & \ttalower\ \ttxmoduleid \\
	& \bnfor & \ttfwrites\ \ttcol\ \sseq{\ttx_i}  \\
	& \bnfor & \ttfmwrites\ \ttcol\ \sseq{\ptrform{\ttmemidi}{\tte_i}}
\end{align*}
\end{minipage}
\hfill\vline\hfill
\begin{minipage}{0.44\textwidth}
\begin{align*}
	& & \textbf{(Initial State Bindings)} \\
	& \ttblockbinds \bnfeq & \eps \bnfor \ttblockbind\ttsemi\ \ttblockbinds \\
	& \ttblockbind \bnfeq & \tLet{\ttx}{\aprimitivetyp}{\ttae}
\end{align*}
\begin{align*}
	& & \textbf{(\ale Specifications)} \\
	& \ttpre \bnfeq & \ttae \\ 
	& \ttpost \bnfeq & \ttae \\
	& \itspec \bnfeq & \ttdecls\ttsemi\ \ttblockbinds\ttsemi\ \ttpre\ttsemi\tsp \ttpost
\end{align*}
\end{minipage}
\caption{\ale declarations, block-lets, and specifications.\label{fig:tr_ale_types2}}
\end{figure*}

\mypara{Declarations and Frames}
\ale declarations come in two forms: \ttareq and \ttaprov, shown in \autoref{fig:tr_ale_types2}.
The second form declares elements in the ordinary way, while
the first form declares an element that must be provided
by the \casp \lowerings or the \casp machine description.
In this case, the type is given, but not the value.
This functions as a form of import, and allows an \ale file to be
checked on its own separately from any particular machine description
or \casp \lowerings.
However, we do not currently define or implement such a check.
Note that it is possible to \ttareq functions that implicitly depend
on machine state or that depend on machine state on some machines and
not others.
Such functions can also depend on constants or other elements that are
not visible in the \ale specification at all.
The \ttalower declarations specify all \lowering modules that 
are used to compile this \ale specification into a \casp specification.
The module name \ttxmoduleid is used to look up the \casp \lowering module to apply.
The \ttaregion declarations declare memory regions, like the
memory-typed \ttletstate declarations in \casp.
(These are implicitly always \ttaprov, because, for memory regions, the
corresponding \ttareq declaration would be entirely equivalent, requiring
duplication in the \casp \lowering.)
Note that the parameters of the region can be symbolic constants if
abstraction is needed.

Frame declarations in \ale, annotated with \ttfwrites and \ttfmwrites, are exactly the same as in \casp.
Because \ale files are machine-independent, the registers mentioned
must be abstract and concretized via the \casp \lowerings.


\mypara{Block-lets}
While \ale expressions include let-bindings, the scope of those
let-bindings is conventional: it lasts until the end of the
expression.
To refer to values taken from the initial state (that is, the
machine state of which the precondition must be true), we need a way to
bind these values so their scope extends to the postcondition.
The block-lets serve this purpose in \ale, shown in \autoref{fig:tr_ale_types2}, 
much like the additional declarations seen in \casp specs can.
These are found within a block (because a block corresponds to a
synthesis problem, it is meaningful to associate before and after
machine states with it), and the scope is the entire block.

\mypara{Specifications}
A full specification starts with a preamble of declarations, shown in \autoref{fig:tr_ale_types2}.
It also includes block-lets and the pre- and post-conditions for the block.
Common declarations can be shared with \ttinclude.

\section{\ale Typing and Semantics}
\label{sec:ale_check}

\begin{figure*}
\textbf{(\casp Integer Constant Extraction)}
\nextrulesmall
\begin{tabular}{ccc}
\irule
	{~}
	{\KdeclAA{\eps}}
&
\irule
	{\KdeclAB{\ttdecl} \nextclause \\
		\KdeclBC{\ttdecls}}
	{\KdeclAC{\ttdecl\ttsemi\ \ttdecls}}
&
\irule
	{\KdeclAB{\ttdecls}}
	{\KmachineB{\ttdecls\ttsemi\ \defops}}
\end{tabular}

\nextrulesmall
\begin{tabular}{ccc}
\irule
	{~}
	{\KdeclAA{\tttype\ \tttypeid = \cbasetyp}}
&
\irule
	{\kenv' = \kenv[\ttx \envgoesto \ttC]}
	{\KdeclAB{\ttlet\ \ttx\ \ttcol\ \ttint = \ttC}}
&
\irule
	{\tte \neq \ttC}
	{\KdeclAA{\ttlet\ \ttx\ \ttcol\ \cbasetyp = \tte}}
\end{tabular}

\begin{tabular}{cc}
\irule
	{~}
	{\KdeclAA{\ttdef\ \ttfuncid\ \funtyp{\sseq{\ttx_i\ \ttcol\ \cbasetypi}}{\cbasetyp} = \tte}}
&
\irule
	{~}
	{\KdeclAA{\ttproc\ \ttprocid\ \proctyp{\sseq{\ttx_i\ \ttcol\ \cbasetypi}} = \ttS}}
\end{tabular}

\begin{tabular}{cc}
\irule
	{~}
	{\KdeclAA{\ttlet\ttstate\ \ttx\ \ttcol\ \cregtyp}}
&
\irule
	{~}
	{\KdeclAA{\ttlet\ttstate\ \ttmemid\ \ttcol\ \cmemtyp}}
\end{tabular}

\begin{tabular}{cc}
\irule
	{~}
	{\KdeclAA{\ttlet\ttstate\ \ttmemid\ \ttcol\ \cmemtyp\ \ttwith\ \NNEW{\ttx}}}
&
\irule
{~}
{\KdeclAA{\ttlet\ \strof{\ttx} = \text{\tte}}}
\end{tabular}

\caption{\casp integer constant extraction.\label{fig:tr_ale_const}}
\end{figure*}

We do not provide (or implement) a full typechecking pass for \ale.
Instead, when we lower to \casp, we allow the \casp typechecker to
reject invalid results, which might be caused by invalid \ale input or by
bad/mismatched \casp \lowering definitions.
The rules provided here are for doing scans over the declarations
sufficient to make the translation to \casp work and no more.

\mypara{Environments}
We retain the \casp typing environments, $\typenv$.
We add an additional environment \kenv, which maps identifiers to
integer constants.
This is a projection of the \casp execution environment $\venv$: it
holds mappings only for variables defined as integer constants and
excludes everything else.
We include a separate set of rules for extracting these integer
constants without doing a full \casp execution.
(Among other things, this avoids involving machine state or the
machine state stores.)

\mypara{Translation}
The translation (\lowering) from \ale to \casp, defined in the next
section, appears cross-recursively in the rules in this section.
Because $\typenv$ are \casp environments, they map identifiers to
\casp types, not \ale ones.
This means \ale types must be lowered on the fly to update
them correctly.

\mypara{Integer Constant Extraction}
The integer constant extraction rules do a simple pass over \casp
declarations to extract the variables defined as integer constants, shown in \autoref{fig:tr_ale_const}.
These populate a substitution environment \kenv that we use for
lowering \ale types containing symbolic constants.
These rules are judgments of the form \KdeclAB{\ttdecl} or
\KdeclAB{\ttdecls}, plus one of the form \KmachineA{\machine} for a
whole machine description.

\begin{figure*}
	\textbf{(\ale Declaration Typing)}
\nextrulesmall
\irule 
{\typenv, \kenv \tholds \ttdecl \ttthen \typenvBB, \kenv' \nextclause \\
	\typenvBB, \kenv \tholds \ttdecls \ttthen \typenvCC, \kenv''}
{\typenv, \kenv  \tholds \ttdecl\ttsemi\ \ttdecls \ttthen \typenvCC, \kenv''}

\nextrulesmall
\begin{tabular}{cc}
\irule 
{	
	\genv \twfholds \tttypeid}
{\typenv, \kenv \tholds \ttareq\ \ttatype\ \tttypeid \ttthen \typenv, \kenv}
&
\irule 
{	\acrule{\aprimitivetyp} = \atyp\nextclause \\
	\tenv(\ttx) = \atyp }
{\typenv, \kenv \tholds \ttareq\ \ttaval\ \hastyp{\ttx}{\aprimitivetyp} \ttthen \typenv, \kenv}
\end{tabular}
\nextrulesmall
\irule 
{   \acrule{\afunctiontyp} = \atyp\nextclause \\
	\tenv (\ttfuncid) = \atyp }
{\typenv, \kenv \tholds \ttareq\ \ttafunc\ \hastyp{\ttfuncid}{\afunctiontyp} \ttthen \typenv, \kenv}
	
\nextrulesmall
\irule 
	{ 
		\acrule{\atyp} = \atyp' \\
		\genv \twfholds \atyp' \nextclause \\
		\genv' = \genv[\tttypeid \goesto \atyp'] }
	{\typenv, \kenv \tholds \ttaprov\tsp\ttatype\tsp\tttypeid = \atyp \ttthen \typenvBA, \kenv }

\nextrulesmall
\irule 
	{ 
		\genv, \env \tholds\hastyp{\ttC}{\ttint} \nextclause \\ 
		\tenv' = \tenv[\ttx  \envgoesto \ttint] \nextclause \\
		\kenv' = \kenv[\ttx  \envgoesto \ttC] }
	{\typenv, \kenv \tholds \ttaprov\tsp\ttaval\tsp\hastyp{\ttx}{\ttint} = \ttC \ttthen \typenvAB, \kenv' }

\nextrulesmall
\irule 
	{ 
		\acrule{\aprimitivetyp} = \atyp\nextclause \\
		\genv \twfholds \atyp \nextclause \\
		\tte \neq \ttC \\
		\acrule{\tte} = \tte' \nextclause \\
		\genv, \env \tholds\hastyp{\tte'}{\atyp} \nextclause \\ 
		\tenv' = \tenv[\ttx  \envgoesto \atyp] }
	{\typenv, \kenv \tholds \ttaprov\tsp\ttaval\tsp\hastyp{\ttx}{\aprimitivetyp} = \ttae \ttthen \typenvAB, \kenv }

\nextrulesmall
\irule 
	{	
		\forall \tti,\ \acrule{\cbasetypi} = {\atyp_i} \wedge \genv \twfholds \atyp_i \nextclause \\
		\acrule{\cbasetyp} = \atyp \nextclause \\
		\genv \twfholds \atyp \nextclause \\ \\
		\acrule{\tte} = \tte' \nextclause \\
		\tenv' = \tenv[\forall \tti, \ttx_i\envgoesto {\atyp_i}] \nextclause \\ 
		\genv, \tenv'\tholds\hastyp{\tte'}{\atyp} \nextclause \\ 
		\tenv'' = \tenv[\ttfuncid \envgoesto \left( \funtyp{\sseq{\ttx_i\ \ttcol\ {\atyp_i}}}{\atyp} \right)] \nextclause \\
		}
	{\typenv, \kenv \tholds \ttaprov\ \ttafunc\ \hastyp{\ttfuncid}{\funtyp{\sseq{\ttx_i\ \ttcol\ \cbasetypi}}{\cbasetyp}} \\
		= \ttae \ttthen \typenvAC, \kenv}

\nextrulesmall
\irule 
	{	
		\acrule{\memtyp{\ttN_1}{\ttN_2}{\ttN_3}} =
			\memtyp{\ttC_1}{\ttC_2}{\ttC_3} \nextclause \\ \\
		\genv \twfholds \memtyp{\ttC_1}{\ttC_2}{\ttC_3} \nextclause \\
		\tenv' = \tenv[\ttmemid  \envgoesto \memtyp{\ttC_1}{\ttC_2}{\ttC_3}]  }
	{\typenv, \kenv \tholds \ttaregion\ \hastyp{\ttmemid}{\memtyp{\ttN_1}{\ttN_2}{\ttN_3}} \ttthen\ \typenvAB, \kenv }

\nextrulesmall
\irule 
 	{   
 		\acrule{\memtyp{\ttN_1}{\ttN_2}{\ttN_3}} =
 			\memtyp{\ttC_1}{\ttC_2}{\ttC_3} \nextclause \\ \\
 		\genv \twfholds \memtyp{\ttC_1}{\ttC_2}{\ttC_3} \nextclause \\
 		\genv \twfholds \labeltyp{\ttC_3} \nextclause \\ \\
 		\acrule{\labeltyp{\ttN_3}} = \labeltyp{\ttC_3} \nextclause \\
		\tenv' = \tenv[\ttmemid  \envgoesto \memtyp{\ttC_1}{\ttC_2}{\ttC_3} \ttsemi\ \NNEW{\ttx} \envgoesto \labeltyp{\ttC_3}] }
	{\typenv, \kenv \tholds \ttaregion\ \hastyp{\ttmemid}{\memtyp{\ttN_1}{\ttN_2}{\ttN_3}}
		\ \ttwith\ \NNEW{\ttx} \ttthen \typenvAB, \kenv }

\irule{~}
	{\typenv, \kenv \tholds \ttalower\ \ttxmoduleid \ttthen \typenv, \kenv}

\nextrulesmall
\irule
{
	\acrule{\ttfwrites\ \ttcol\ \sseq{\ttx_i}} =
		\ttfwrites\ \ttcol\ \sseq{\ttx_i}
	\nextclause \\
	\typenv \tholds
		\ttfwrites\ \ttcol\ \sseq{\ttx_i}
}
{
	\typenv, \kenv \tholds
		\ttfwrites\ \ttcol\ \sseq{\ttx_i}
		\ttthen \typenv, \kenv
}

\nextrulesmall
\irule
{
	\acrule{\ttfmwrites\ \ttcol\ \sseq{\ptrform{\ttmemidi}{\tte_i}}} =
		\ttfmwrites\ \ttcol\ \sseq{\ptrform{\ttmemidi}{\tte_i'}}
	\nextclause \\
	\typenv \tholds
		\ttfmwrites\ \ttcol\ \sseq{\ptrform{\ttmemidi}{\tte_i'}}
}
{
	\typenv, \kenv \tholds
		\ttfmwrites\ \ttcol\ \sseq{\ptrform{\ttmemidi}{\tte_i}}
		\ttthen \typenv, \kenv
}

\caption{\ale typing rules for declaration.\label{fig:tr_ale_type_decl}}
\end{figure*}

\mypara{Typing}
The declaration typing rules are intended to accumulate types for all
the declarations in an \ale specification.
They are applied concurrently with the \casp declaration rules to the
\ale specification, the \casp machine description, and the \casp
\lowering.
The declaration typing rules have judgments of the form
$\typenv, \kenv \tholds \ttdecl \ttthen \typenvBB, \kenv'$ and
$\typenv, \kenv \tholds \ttdecls \ttthen \typenvBB, \kenv'$, shown in \autoref{fig:tr_ale_type_decl}.
These mean that the declaration or declarations update the type
environment (and integer constant environment) as shown.
Note that there is a special-case rule for \ttaprov \ttaval for when
the value is an integer constant; this enters the constant into \kenv.
The integer constants are in turn used when lowering the types of
memory regions, which can be seen in the last two rules.

\begin{figure*}
	\textbf{(\ale Specification Typing)}

\nextrulesmall
\begin{tabular}{cc}
\irule 
{~}
{\typenv, \kenv \tholds \eps \ttthen \typenv, \kenv}
&
\irule 
	{\typenv, \kenv \tholds \ttblockbind \ttthen \typenvBB, \kenv' \ \ \ \ \ 
		\typenvBB, \kenv  \tholds \ttblockbinds \ttthen \typenvCC, \kenv'' }
	{\typenv, \kenv \tholds \ttblockbind\ttsemi\ \ttblockbinds \ttthen \typenvCC, \kenv'' }
\end{tabular}

\nextrulesmall
\irule 
{	
	\acrule{\cbasetyp} = \atyp \ \ \ \ \ 
	\genv \twfholds \atyp  \ \ \ \ \ 
	\acrule{\tte} = \tte' \ \ \ \ \ 
	\genv, \env \tholds\hastyp{\tte'}{\atyp} \ \ \ \ \ 
	\tenv' = \tenv[\ttx \envgoesto \atyp] }
{\typenv, \kenv \tholds \tLet{\ttx}{\aprimitivetyp}{\ttae}  \ttthen \typenvAB, \kenv }

\nextrulesmall
\irule
{
	\typenv, \kenv \tholds \ttdecls \ttthen \typenvBB, \kenv'
		\nextclause \\
	\typenvBB, \kenv' \tholds \ttblockbinds \ttthen  \typenvCC, \kenv''
		\nextclause \\
	\acruleCC{\ttpre} = \ttpre'
		\nextclause \\
	\acruleCC{\ttpost} = \ttpost'
		\nextclause \\
	\typenvCC \tholds \hastyp{\ttpre'}{\ttbool}
		\nextclause \\
	\typenvCC \tholds \hastyp{\ttpost'}{\ttbool}
}
{
	\typenv, \kenv \tholds
	    \ttdecls\ttsemi\ \ttblockbinds\ttsemi\ \ttpre\ttsemi\tsp \ttpost
	\ttthen \typenvCC, \kenv''
}

\caption{\ale typing rules for specifications.\label{fig:tr_ale_type_spec}}
\end{figure*}

\mypara{Block-lets}
The rules for block-lets are effectively the same as the rules for
declarations, shown in \autoref{fig:tr_ale_type_spec}.
The ways in which block-lets are special mostly do not apply here.
Note however that even though we pass through \kenv (for consistency
of the form of the rules) there is no rule for loading integer
constants into \kenv from block-lets.
Integer constants used in types and defined in the \ale specification should
be defined with \ttaprov \ttaval; block-lets are intended to provide
access to machine state.

\begin{figure*}
	\textbf{(\ale Specification Semantics)}
\nextrulesmall
\irule 
	{
		\itspec = 
			\ttdecls_{\mathit{ale}}\ttsemi\
			\ttblockbinds\ttsemi
			\ \ttpre\ttsemi\tsp \ttpost
		\nextclause \\\\
		\tholds \machine \ttthen \genv_0, \tenv_0 \ \ \ \ \ \ 
		\KmachineZ{\machine} \ \ \ \ \ \ 
		\left( \genv_0 \subseteq \genv \right) \wedge \left( \tenv_0 \subseteq \tenv \right) \wedge \left( \kenv_0 \subseteq \kenv \right)\nextclause \\
		\genv, \tenv, \kenv \tholds \ttdecls_{\mathit{ale}} \ttthen \genv, \tenv, \kenv \ \ \ \ \ \ 
		\acrule{\ttdecls_{\mathit{ale}}} \subseteq \ttdecls\ttsemi\ \ttframes \nextclause \\
		( \forall \ttxmoduleid,
			\ttalower\ \ttxmoduleid \in \ttdecls_{\mathit{ale}}
		\implies \nextclause \\
			\ttmodule\ \ttxmoduleid
			\ \{\ \ttdecls_{\mathit{lower}}\ttsemi
				\ \ttframes_{\mathit{lower}}\ \}
			\in \itmodules
		\wedge
			\ttdecls_{\mathit{lower}} \subseteq \ttdecls
		\wedge
			\ttframes_{\mathit{lower}} \subseteq \ttframes
		) \nextclause \\
		\genv, \tenv \tholds \decls \ttthen \genv, \tenv \ \ \ \ \ \ 
		\kenv \tholds \decls \ttthen \kenv \nextclause \\
		\genv, \tenv, \kenv \tholds \ttblockbinds \ttthen \genv, \tenv, \kenv \ \ \ \ \ \ 
		\acrule{\ttblockbinds} \subseteq \ttdecls \nextclause \\
		\acrule{\ttpre} = \ttpre' \nextclause \\
		\acrule{\ttpost} = \ttpost' \nextclause \\
		\Omega = \ttdecls\ttsemi\ \ttframes\ttsemi\ \ttpre'\ttsemi\tsp \ttpost'
	}
	{\machine, \itmodules, \itspec \ttthen \Omega}
\caption{\ale semantics for specifications.\label{fig:tr_ale_sem_spec}}
\end{figure*}

\mypara{Specifications}
The rule for the semantics of an entire specification is large and
complex.
The conclusion is that a given machine, \lowering module, and \ale
specification produce a final translation output $\Omega$.
The rules work by nondeterministically
taking fixpoints over all the material included.
$\ttdecls$ is the combination of all declarations found both in the
initial specification and all the included lowering modules, and
$\ttframes$ is the combination of all frame information (part of
the declarations in \ale; separated in \casp).
Similarly, the final set of environments $\tenv, \genv, \kenv$
represent fixpoints produced by processing all the declarations.

In \autoref{fig:tr_ale_sem_spec}, 
the first premise expands the \ale
specification as we will need to work with the components.
The next two premises generate initial environments: the \casp typing
environments induced by the machine description, and its integer
constants, and then we require that these are included in the
final environments.
In the fifth and sixth premises, we then require that 
the result of processing the declarations from
the specification appears in the final environments, and that the
translation of these to \casp is included in the final declarations
and frame rules.
Then for every lowering module requested by the specification, we
require that it be provided in the input modules list and that its
components appear in the final declarations and frame rules.
This is followed by two more rules to ensure that these results
are represented in the final environments.
Later, we include the block-let material in the final environments,
include its lowered form in the final declaration list (block-lets
lower to declarations), bind
the lowerings of the pre- and postconditions, and define
the output.

The fixpoint-based evaluation strategy for declarations is required,
because the
\ale declarations rely on the \casp \lowering file (most notably for
resolving symbolic constants), but the \casp \lowering file is in turn
also specifically allowed to refer to objects declared by the \ale
specification, such as memory regions.
In the implementation this circularity is resolved by lifting both the
\casp and \ale declarations (and block-lets) into a common
representation and topologically sorting them based on identifier
references.
(Genuinely circular references among identifiers are prohibited.)
From this point, they can be handled in order in a more conventional
way.

\mypara{Complete Output}
Note that the output includes the declarations
from the \casp \lowering modules (each $\ttdecls_{\mathit{lower}}$).
Apart from symbolic constants, we do not substitute the definitions of
the \lowering elements,
as that would greatly complicate things,
especially with functions; instead we include the definitions and let
the translation refer to them.
In fact, because of the declaration ordering issues, in the
implementation the \lowering declarations and
translated \ale declarations can be arbitrarily interleaved in the
output.

Note furthermore that it would not be sufficient to include
\emph{only} the \lowering declarations explicitly imported with \ttareq
declarations, as those may refer freely to other things declared in
the \lowering module that the \ale specification itself may have no cognizance
of whatsoever.

\newpage

\section{Lowering \ale}
\label{sec:ale_exec}
\begin{figure*}
	\textbf{\ale{} -- \casp{} Type Translation}
	\begin{align*}
\actyprule{\ttN} = &
	\begin{cases} 
	\ttC & \ttN = \ttC \\
	\acsigma{\ttx} & \ttN = \ttx \wedge \ttx \in \kenv \\
	\bot & \ttN = \ttx \wedge \ttx \notin \kenv\\
	\end{cases}
&
\actyprule{\tttypeid} = &
	\begin{cases} 
		\acdelta{\tttypeid} & \tttypeid \in \genv \\
		\bot & \tttypeid \notin \genv
		\end{cases} \\
\actyptrans{\ttint} = &~ \ttint
&
\actyptrans{\ttbool} = &~ \ttbool \\
\actyptrans{\ttN\ \ttwvec} = &~ \bvaltyp{ \actyptrans{\ttN} } &
\actyptrans{\ttN\ \ttwptr} = &~ \bvaltyp{ \actyptrans{\ttN} }\\
\actyptrans{\ttN\ \abvltyp} = &~ \bvltyp{ \actyptrans{\ttN} } &
\actyptrans{\ttN\ \abvstyp} = &~ \bvstyp{ \actyptrans{\ttN} }
\end{align*}
\begin{align*}
\actyptrans{\memtyp{\ttN_1}{\ttN_2}{\ttN_3}} = &~ \ensm{\actyptrans{\ttN_1}\ \ttbit}\ \ensm{\actyptrans{\ttN_2} \ \ttmlen\ \actyptrans{\ttN_3}\ \texttt{ref}} \\
\actyptrans{\afuntyp} = &~ \funtyp{\sseq{\actyptrans{\cbasetypi}}}{\actyptrans{\cbasetyp}}
\end{align*}
\caption{\ale{} -- \casp{} type translation.\label{fig:tr_lower_type}}
\end{figure*}

The semantics of an \ale specification depend on material taken
from a \casp mapping and machine description.
This does not preclude defining a semantics for \ale in terms of
that material or even some abstracted concept of what any such \casp
material might be.
However, doing so is complicated (as can be seen from the material in the
previous section, which does not even attempt to handle expression
evaluation) and not necessarily rewarding or illuminating.

So instead, we write only enough typing
rules to prepare material for writing a translation
(lowering) to \casp, and then apply the \casp typing (and, implicitly,
semantics) to the
lowered material.
That material goes into the \casp typing environments $\typenv$, and as
discussed in the previous section, we also maintain an additional
environment \kenv of integer constants used for substituting symbolic
constants in types.

This section defines the translation.
$\actrans{a}$ defines the \casp lowering of an \ale element $a$.
We make the translation polymorphic over the various kinds of element;
that is, $\actrans{\atyp}$ is the translation of a type (shown in \autoref{fig:tr_lower_type}),
$\actrans{\tte}$ is the translation of an expression, etc.
Some of the translation rules rely on the environments; these are
written $\acrule{a}$ (shown in \autoref{fig:tr_lower_all}).

Some of the translation rules produce $\bot$.
If these are reached, the translation fails; this can happen if the
\ale spec was malformed and, potentially, if the mapping module
failed to declare elements that were expected of it or declared them
in an incompatible or inconsistent way.
The rules in the previous section exclude some of these cases, but we
are not (yet) prepared to argue that they rule out all
translation-time failures.

Notice that the translations for \ttareq declarations are empty,
because the declarations from the mapping module are output along with
the translated \ale specification.

\begin{figure*}
\textbf{\ale{} -- \casp{} Expression Translation}
\begin{align*}
\acrule{\ttx} = &
	\begin{cases}
 	\ttx & \ttx \in \tenv \\
	\bot & \ttx \notin \tenv
 	\end{cases} 
&
\acrule{\fapp{\ttfuncid}{\sseq{\ttae}}} = &~
\begin{cases} 
	\fapp{\ttfuncid}{\sseq{\actrans{\ttae}}} & \ttfuncid \in \tenv \\
	\bot & \ttfuncid \notin \tenv
\end{cases} \\
\actrans{\tttrue} = &~ \tttrue &
\actrans{\ttfalse} = &~ \ttfalse \\
\actrans{\ttC} = &~ \ttC &
\actrans{\bitconst} = &~ \bitconst \\
\actrans{\itunop\ \ttae} = &~ \itunop\ \actrans{\ttae} &
\actrans{\ttae_1\ \itbinop\ \ttae_2} = &~ \actrans{\ttae_1}\ \itbinop\ \actrans{\ttae_2} \\
\actrans{\bitidx{\ttae}{\ttC}} = &~ \bitidx{\actrans{\ttae}}{\ttC} &
\actrans{\bitfield{\ttae}{\ttC_1}{\ttC_2}} = &~ \bitfield{\actrans{\ttae}}{\ttC_1}{\ttC_2}
\end{align*}
\vspace{-.6cm}
\begin{align*}
\actrans{\eLet{\ttx}{\cbasetyp}{\ttae_1}{\ttae_2}} = &~ \ensm{\texttt{let}\tsp\oftyp{\ttx}{\actyptrans{\cbasetyp}} } \ensm{ = \actrans{\ttae_1}\ \ensm{\texttt{in}\tsp{\actrans{\ttae_2}}}} \\
\actrans{\sITE{\ttae_1}{\ttae_2}{\ttae_3}} = &~ \ensm{\texttt{if}\tsp{\actrans{\ttae_1}}}\ \ensm{\texttt{then}\tsp{\actrans{\ttae_2}}}\ \ensm{\ensm{\texttt{else}\tsp{\actrans{\ttae_3}}}} \\
\acrule{\ptrform{\ttmemid}{\ttae}} = &
	\begin{cases} 
	\ptrform{\ttmemid}{\actrans{\ttae}} & \ttmemid \in \tenv \\
	\bot & \ttmemid \notin \tenv
	\end{cases}
\end{align*}
\vspace{-.3cm}
\begin{align*}
\actrans{\ttassertfalse} = &~ \ttassertfalse \\
\actrans{\regderef{\ttae}} = &~ \regderef{\actrans{\ttae}} &
\actrans{\fetch{\ttae}{\ttN}} = &~ \fetch{\actrans{\ttae}}{\actyptrans{\ttN}} \\
\NEW{\actrans{\ebranchto{sym}} = }&~\NEW{ \ebranchto{sym}} &
\actrans{\setlit} = &~ \ensm{\ttinbrace{\actrans{\ttx_1}, \ldots, \actrans{\ttx_k}}} \\
\actrans{\sizeofset{\ttae}} = &~ \sizeofset{\actrans{\ttae}} &
\actrans{\setmemberof{\ttae_1}{\ttae_2}} = &~ \setmemberof{\actrans{\ttae_1}}{\actrans{\ttae_2}}
\end{align*}
\vspace{-.5cm}
\nextrulesmall
\textbf{\ale{} -- \casp{} Block-Lets Translation}
\begin{align*}
	\actrans{\tLet{\ttx}{\aprimitivetyp}{\ttae}} = &~ \tLet{\ttx}{\actrans{\aprimitivetyp}}{\actrans{\ttae}} 
\end{align*}
\vspace{-.5cm}
\nextrulesmall
\textbf{\ale{} -- \casp{} Declaration Translation}
\begin{align*}
\acrule{\ttareq\ \ttatype\ \tttypeid} = &
	\begin{cases} 
		\eps & \tttypeid \in \genv \\
		\bot & \tttypeid \notin \genv
		\end{cases} \\
\acrule{\ttareq\ \ttaval\ \hastyp{\ttx}{\aprimitivetyp}} = &
	\begin{cases} 
	\eps & \ttx \in \tenv \\
	\bot & \ttx \notin \tenv
	\end{cases} \\
\acrule{\ttareq\ \ttafunc\ \hastyp{\ttfuncid}{\afunctiontyp} } = &
	\begin{cases} 
	\eps & \ttfuncid \in \tenv \\
	\bot & \ttfuncid \notin \Sigma
	\end{cases} \\
\actrans{\ttaprov\tsp\ttatype\tsp\tttypeid = \atyp} = &~ \tttype\ \tttypeid = \actyptrans{\atyp}\\
\actrans{\ttaprov\tsp\ttaval\tsp\hastyp{\ttx}{\aprimitivetyp} = \ttae} = &~ \ttlet\ \ttx\ \ttcol\ \actyptrans{\cbasetyp} = \actrans{\tte}\\
\actrans{\ttaprov\ \ttafunc\ \hastyp{\ttfuncid}{\funtyp{\sseq{\ttx_i\ \ttcol\ \cbasetypi}}{\cbasetyp}} = \ttae} = &~ \ttdef\ \ttfuncid\ \funtyp{\sseq{\ttx_i\ \ttcol\ \actyptrans{\cbasetypi}}}{\actyptrans{\cbasetyp}} \\
	& ~ = \actrans{\tte}\\
\actrans{\ttaregion\ \hastyp{\ttmemid}{\amemorytyp}} = &~ \ttlet\ttstate\ \ttmemid\ \ttcol\ \actyptrans{\cmemtyp}\\
\actrans{\ttaregion\ \hastyp{\ttmemid}{\amemorytyp}\ \ttwith\ \NNEW{\ttx}} = &~ \ttlet\ttstate\ \ttmemid\ \ttcol\ \actyptrans{\cmemtyp}\ \ttwith\ \NNEW{\ttx} \\
\actrans{\ttalower\ \ttxmoduleid} = &~ \epsilon \\
\actrans{\ttfwrites\ \ttcol\ \sseq{\ttx_i}} = &~ \ttfwrites\ \ttcol\ \sseq{\actrans{\ttx_i}} \\
\actrans{\ttfmwrites\ \ttcol\ \sseq{\ptrform{\ttmemidi}{\tte_i}}} = &~ \ttfmwrites\ \ttcol\ \sseq{\ptrform{\actrans{\ttmemidi}}{\actrans{\tte_i}} }
\end{align*}
\caption{\ale{} -- \casp translation.\label{fig:tr_lower_all}}
\end{figure*}

\newpage

\section{Conclusion}
In this technical report, we described two domain
specific languages involved in the \aquarium kernel synthesis
project.
First, we presented a machine modeling language named
\casp that can be used used to describe the semantics of many
different processor ISAs at the assembly language level.
Then, we presented a specification language named \ale that allows
stating abstract specifications for blocks of assembly code, such that
these abstract specifications can be lowered to concrete
specifications and used for synthesis and verification against
\casp machine descriptions.

We note that this is work in progress, and does not yet present a
final or complete view of either the \aquarium system or the languages
presented.
\balance

\bibliography{ms}

\end{document}